# The Cold Molecular Gas Regulates the Activity of Active Galactic Nuclei in Massive Galaxies

Yongyun Chen (陈永云)[1], Qiusheng Gu (顾秋生)[2], Luis.C Ho (何子山)[3], Junhui Fan (樊军辉)[4], Feng Yuan (袁峰)[5], Tao Wang (王涛)[2], Zhifu Chen (陈志福)[6], Dingrong Xiong (熊定荣)[7], Xiaoling Yu (俞效龄)[1], Xiaotong Guo (郭晓通)[8], and Nan Ding (丁楠)[9]

[1] College of Physics and Electronic Engineering, Qujing Normal University, Qujing 655011, People's Republic of China; ynkmcyy@yeah.net
[2] School of Astronomy and Space Science, Nanjing University, Nanjing 210093, People's Republic of China; qsgu@nju.edu.cn
[3] Kavli Institute for Astronomy and Astrophysics, Peking University, Beijing 100871, People's Republic of China
[4] Center for Astrophysics, Guang zhou University, Guang zhou510006, People's Republic of China
[5] Center for Astronomy and Astrophysics and Department of Physics, Fudan University, Shanghai 200438, People's Republic of China
[6] School of Mathematics and Physics, Guangxi Minzu University, Nanning 530006, People's Republic of China
[7] Yunnan Observatories, Chinese Academy of Sciences, Kunming 650011, People's Republic of China
[8] Anqing Normal University, 246133, People's Republic of China
[9] School of Physical Science and Technology, Kunming University 650214, People's Republic of China

## Abstract

The physical quantities that directly regulate active galactic nucleus (AGN) feedback in massive galaxies remain poorly understood. Observations of molecular gas surrounding AGNs suggest that this gas serves as a fuel source for AGN activity. Accordingly, we study the relationship between AGN activity and molecular gas properties. In this study, we analyze a large sample of nearby AGNs with available measurements of molecular gas mass, radio luminosity, and [O III] luminosity. Our results show that radio luminosity and [O III] luminosity exhibit stronger correlations with molecular gas mass than with other physical parameters such as black hole mass, stellar mass, and bulge mass. Moreover, when controlling for the correlations between radio luminosity, [O III] luminosity, and molecular gas mass, the relationships between these luminosities and other key physical parameters become significantly weaker or disappear entirely. This suggests that, of all the properties we have considered, it is the molecular gas mass that is most tightly correlated with radio and [O III] luminosity, and may thus be the most important driver of nuclear activity.

## 1. Introduction

The evolution of galaxies is closely linked to the growth of their supermassive black holes, as demonstrated by the observed correlations between black hole mass and various properties of the host galaxy (e.g., J. Magorrian et al. 1998; L. Ferrarese & D. Merritt 2000; K. Gebhardt et al. 2000; J. Kormendy & R. C. Kennicutt 2004; J. Kormendy & L. C. Ho 2013). The connection between black hole accretion and host galaxy growth arises naturally, as both processes depend on a shared gas reservoir, which is primarily supplied by similar mechanisms that drive gas inward (J. Kormendy & R. C. Kennicutt 2004). The question of whether, when, and how supermassive black holes coevolve with their host galaxies has long been a topic of debate (J. Kormendy & L. C. Ho 2013; T. M. Heckman & P. N. Best 2014; J. E. Greene et al. 2020). It is widely believed that this coevolution is regulated by feedback from active galactic nuclei (AGNs; A. C. Fabian 2012). During black hole accretion, a substantial amount of energy is released, although only a small fraction is effectively coupled to the surrounding environment, where it can heat and/or expel gas from the host galaxy (R. S. Somerville et al. 2008; J. Schaye et al. 2015; D. Nelson et al. 2018), thereby suppressing ongoing star formation (Y. Dubois et al. 2016). AGN feedback can also influence the galaxy halo by preventing the condensation of cold and warm gas, thus inhibiting star formation over extended timescales (R. G. Bower et al. 2006; D. J. Croton et al. 2006; A. C. Fabian 2012; M. Gaspari et al. 2020).

Supermassive black holes are ubiquitous in the centers of galaxies (J. Kormendy & L. C. Ho 2013), and they generate huge amounts of energy as AGNs. AGNs in massive elliptical galaxies typically operate in the "radio feedback" mode and frequently exhibit jet activity (T. M. Heckman & P. N. Best 2014). Given that elliptical galaxies are filled with hot gas (G. Fabbiano et al. 1989), it is reasonable to assume that this gas accretes onto black holes via Bondi accretion (H. Bondi 1952). S. W. Allen et al. (2006) reported a strong correlation between the X-ray cavity jet power ($P_{\rm cav}$) of AGNs and the Bondi accretion rate. In contrast, H. R. Russell et al. (2013) found only a weak correlation between these two quantities. A significant amount of molecular gas has been detected in elliptical galaxies at the centers of nearby galaxy clusters (A. C. Edge 2001; P. Salomé & F. Combes 2003; L. P. David et al. 2014; B. R. McNamara et al. 2014; H. R. Russell et al. 2016, 2017, 2019; V. Olivares et al. 2019; T. Rose et al. 2020; G. Schellenberger et al. 2020; E. V. North et al. 2021), suggesting that this cold gas may serve as a fuel source for AGNs. Numerous studies have identified absorption lines in certain AGNs, which provide direct evidence for AGN fueling and indicate the presence of dense gas near the central regions (L. P. David et al. 2014; G. R. Tremblay et al. 2016; T. Rose et al. 2019; T. Rose et al. 2023). R. D. Baldi et al. (2015) found a weak correlation between [O III] luminosity and molecular gas mass in a sample of 37 early-type galaxies. H. R. Russell et al. (2019) observed a strong correlation between X-ray cavity jet power and molecular gas mass (see their Figure 7). Similarly, Y. Fujita et al. (2024) reported a strong correlation using a sample of 22 elliptical galaxies. However, some studies have found no significant dependence of AGN activity on molecular gas properties (e.g., J. Molina et al. 2023b). In addition to molecular gas mass, several authors have suggested that AGN activity is also influenced by black hole mass (e.g., P. N. Best et al. 2005; Y. Liu et al. 2006; J. Sabater et al. 2019; G. Soares & R. Nemmen 2020; Y. Chen et al. 2023). Notably, Y. Fujita et al. (2024) found no correlation between X-ray cavity jet power and black hole mass in their sample of 22 elliptical galaxies. Therefore, the question remains open: does AGN activity primarily depend on molecular gas mass or black hole mass?

Over the past decade, numerous studies have aimed to determine whether AGN feedback can effectively remove sufficient amounts of cold gas from host galaxies to suppress ongoing star formation. Whether and how luminous AGNs can remove and/or heat gas from their host galaxies, thereby influencing star formation activity, remain uncertain (C. M. Harrison 2017). It is also still debated whether AGN activity is associated with the presence of molecular gas. Optically visible and largely unobscured quasars—the most luminous type of AGNs—are ideal targets for investigating the potential impact of AGN feedback on host galaxies. Luminous AGNs are capable of driving multiphase outflows and transferring energy into the interstellar medium (ISM) of their host galaxies (e.g., C. Cicone et al. 2014; R. Morganti 2017; S. Veilleux et al. 2020; A. Girdhar et al. 2022; J. Molina et al. 2023b). However, direct observational evidence of black hole impacts on the galaxy ISM has so far been limited to a few extreme cases (L. Bîrzan et al. 2004; K. W. Cavagnolo et al. 2010). For the general galaxy population, it remains unclear whether and how black holes influence the ISM. We also note that previous studies examining the relationship between AGN power and molecular gas mass have often been based on small sample sizes, which may lead to spurious or misleading correlations. In the context of galaxy formation modeling, various gas accretion models onto black holes have been adopted. For instance, the Bondi accretion model is commonly used in Illustris simulations (e.g., D. Sijacki et al. 2015), while some semianalytical models rely on empirical prescriptions for cold gas accretion (e.g., M. Enoki et al. 2014). Therefore, the observed correlation between AGN power and molecular gas mass could potentially serve as an alternative to these more complex models.

In this study, we employ radio luminosity and [O III] luminosity as tracers of AGN activity to investigate the relationship between AGN activity and molecular gas mass using a large sample of AGNs. We also examine whether black hole mass plays a key role in regulating AGN activity. This paper is organized as follows: Section 2 describes the sample selection. Section 3 presents the results and discusses their implications. Section 4 summarizes the main conclusions. A $\Lambda$CDM cosmology with $H_0 = 70\ \mathrm{km\ s^{-1}\ Mpc^{-1}}$, $\Omega_\Lambda = 0.73$, and $\Omega_\mathrm{m} = 0.27$ is adopted throughout the analysis.

## 2. The Sample

We have assembled a comprehensive sample of AGNs with measurements of molecular gas mass, stellar mass, star formation rates (SFRs), and black hole mass. Specifically, we focus on the host galaxies of AGNs that provide data on stellar mass, molecular gas mass, and SFRs (A. Saintonge et al. 2017; L. Lin et al. 2020; J. Shangguan et al. 2020a; M. J. Koss et al. 2021; I. Smirnova-Pinchukova et al. 2022; D. Wylezalek et al. 2022; S. J. Molyneux et al. 2024). J. Shangguan et al. (2020a) presented data for 40 Palomar–Green quasars, including molecular gas mass, stellar mass, and SFRs. M. J. Koss et al. (2021) provided data for 213 nearby Swift's Burst Alert Telescope (BAT) AGNs, also covering molecular gas mass, stellar mass, and SFRs. I. Smirnova-Pinchukova et al. (2022) reported on 41 AGNs from the Close AGN Reference Survey (CARS), while S. J. Molyneux et al. (2024) detailed 17 quasars from the Quasar Feedback Survey (QFS). Additionally, we collect data on molecular gas mass, stellar mass, and SFRs from the xCOLD GASS survey (A. Saintonge et al. 2017) for xGASS galaxies, as well as from the Arizona Radio Observatory Survey (D. Wylezalek et al. 2022) and Atacama Large Millimeter Array survey for MaNGA galaxies (L. Lin et al. 2020). We only include xGASS and MaNGA galaxies classified as AGNs (J. M. Comerford et al. 2020). Furthermore, we consider the morphological parameters of AGN host galaxies (D. Makarov et al. 2014). In total, our final sample comprises 411 AGNs with available measurements of stellar mass, molecular gas mass, and SFRs (see Table 1). The stellar masses of these AGNs range from $10^{9.5} M_\odot$ to $10^{11.5} M_\odot$.

### 2.1. The Black Hole Mass

The black hole mass measurements for BAT-AGNs are obtained from the Swift-BAT AGN Spectroscopic Survey DR2 (M. J. Koss et al. 2022). For Palomar–Green quasars, the black hole masses are estimated using the broad H$\beta$ full width at half-maximum (FWHM) and the 5100Å continuum luminosity (J. Shangguan et al. 2020a). The black hole masses of the CARS sample are determined using the broad H$\beta$ FWHM and H$\beta$ luminosity (B. Husemann et al. 2022). The black hole masses of the QFS sample are taken from J. R. Mullaney et al. (2013). For the xGASS and MaNGA samples, we derived the black hole masses using their velocity dispersion ($\sigma_\mathrm{SDSS}$) from the NASA-Sloan Atlas catalog[10] (R. C. E. van den Bosch 2016; $\sigma_\mathrm{SDSS}$, and we require $\sigma_\mathrm{SDSS} \geqslant 70\ \mathrm{km\ s^{-1}}$):

$$\log \frac{M_\mathrm{BH}}{M_\odot} = (8.32 \pm 0.04) + (5.35 \pm 0.23) \log\left(\frac{\sigma_\mathrm{SDSS}}{200\ \mathrm{km\ s^{-1}}}\right). \quad (1)$$

The uncertainty of black hole mass estimated from the $M_\mathrm{BH}$–$\sigma$ relation is 0.3 dex and from the virial method is 0.5 dex (F. Ricci et al. 2022). The black hole mass of our sample mainly comes from the above two methods. Therefore, we use an average uncertainty of 0.4 dex.

[10] http://www.nsatlas.org

**Table 1**
The Sample of Massive Galaxies

| Name (1) | R.A. (2) | Decl. (3) | $z_*$ (4) | T-type (5) | $\log M_{BH}/M_\odot$ (6) | $\log M_{H_2}/M_\odot$ (7) | $\log M_*/M_\odot$ (8) | $\log L_{[O III]}$ (9) | $\log L_{[1.4 GHz]}$ (10) | $\log L_{[3 GHz]}$ (11) | log SFR (12) |
|---|---|---|---|---|---|---|---|---|---|---|---|
| 2MASXJ00004876-0709117 | 0.2032 | −7.1532 | 0.037 | 0.1 | 7.61 | 9.08 | $10^{+0.3}_{-0.3}$ | 40.639 ± 0.004 | ... | ... | $0.79^{+0.3}_{-0.3}$ |
| NGC7811 | 0.6101 | 3.3519 | 0.025 | 5 | 6.7 | 9.19 | $10.64^{+0.3}_{-0.3}$ | 40.641 ± 0.001 | 38.07 ± 0.04 | ... | $0.6^{+0.07}_{-0.07}$ |
| Mrk335 | 1.5814 | 20.203 | 0.026 | −5 | 7.23 | 9.85 | $10.75^{+0.3}_{-0.3}$ | 41.36 ± 0.001 | 38.2 ± 0.03 | 38.35 ± 0.03 | $-0.35^{+0.17}_{-0.17}$ |
| PG0007+106 | 2.629191 | 10.974862 | 0.089 | −1.1 | 8.87 | 9.15 | $10.84^{+0.3}_{-0.3}$ | 42.47 ± 0.016 | 38.9 ± 0.02 | 40.45 ± 0 | $0.75^{+0.02}_{-0.03}$ |
| SDSS J001934.54+161215.0 | 4.894008392 | 16.20419338 | 0.037 | −3 | 7.67 | 8.51 | $10.84^{+0.38}_{-0.53}$ | 39.554 ± 0.042 | ... | ... | $-1.43^{+0.74}_{-0.09}$ |
| SDSS J002558.89+135545.8 | 6.495370784 | 13.92936253 | 0.042 | −5 | 7.86 | 8.54 | $10.69^{+0.38}_{-0.51}$ | 39.718 ± 0.039 | 38.81 ± 0.01 | ... | $-1.34^{+0.74}_{-0.08}$ |
| ESO112-6 | 7.6826 | −59.0072 | 0.029 | 2.4 | 7.9 | 9.52 | $10.56^{+0.3}_{-0.3}$ | 39.768 ± 0.033 | ... | ... | $0.79^{+0.3}_{-0.3}$ |
| Mrk344 | 9.6338 | 23.6134 | 0.025 | 0 | 7.54 | 8.73 | $10.48^{+0.3}_{-0.3}$ | 40.33 ± 0.001 | 39.6 ± 0.02 | 37.73 ± 0.12 | $-0.14^{+0.11}_{-0.11}$ |
| SDSS J003921.66+142811.5 | 9.840248599 | 14.46989069 | 0.038 | −5 | 7.09 | 8.52 | $10.37^{+0.41}_{-0.48}$ | 39.477 ± 0.047 | ... | ... | $-0.9^{+0.68}_{-0.11}$ |
| HE0040-1105 | 10.653585 | −10.822785 | 0.042 | −2 | 6.74 | 8.19 | $10.16^{+0.13}_{-0.1}$ | 41.314 ± 0 | ... | 37.98 ± 0.14 | $0.51^{+0}_{0}$ |
| NGC235A | 10.72 | −23.541 | 0.022 | −2.2 | 8.49 | 9.7 | $10.9^{+0.3}_{-0.3}$ | 41.131 ± 0 | 39.22 ± 0.02 | 38.84 ± 0.01 | $0.81^{+0.07}_{-0.07}$ |
| MCG-2-2-95 | 10.7866 | −11.601 | 0.019 | −1 | 6.82 | 8.49 | $10.03^{+0.3}_{-0.3}$ | 40.049 ± 0.015 | 38.41 ± 0.07 | ... | $-0.82^{+0.05}_{-0.05}$ |
| 2MASXJ01003490-4752033 | 15.1457 | −47.8677 | 0.048 | −0.1 | 8.06 | 10.03 | $10.95^{+0.3}_{-0.3}$ | 41.559 ± 0.022 | 37.93 ± 0.01 | ... | $0.8^{+0.08}_{-0.08}$ |
| MCG-7-3-7 | 16.3617 | −42.2162 | 0.03 | 1 | 8.39 | 8.81 | $10.87^{+0.3}_{-0.3}$ | 41.235 ± 0.002 | 38.98 ± 0.02 | ... | $0.21^{+0.3}_{-0.3}$ |
| ESO243-26 | 16.4083 | −47.0717 | 0.019 | 4.1 | 7.24 | 9.1 | $10.39^{+0.3}_{-0.3}$ | 39.265 ± 0.091 | 38.18 ± 0.02 | ... | $0.27^{+0.03}_{-0.03}$ |
| UM85 | 16.6886 | 6.6338 | 0.041 | −2 | 7.82 | 9.22 | $10.64^{+0.3}_{-0.3}$ | 41.811 ± 0.003 | 38.36 ± 0.01 | 38.47 ± 0.05 | $-0.07^{+0}_{0}$ |
| 2MASXJ01073963-1139117 | 16.9152 | −11.6531 | 0.047 | −1 | 7.96 | 9.49 | $10.91^{+0.3}_{-0.3}$ | 41.296 ± 0.001 | 38.19 ± 0.03 | 38.93 ± 0.06 | $0.8^{+0.3}_{-0.3}$ |
| SDSS J010905.96+144520.8 | 17.27485573 | 14.75578944 | 0.039 | −5 | 6.72 | 8.5 | $10.35^{+0.42}_{-0.46}$ | 39.097 ± 0.075 | 38.79 ± 0.02 | ... | $-0.91^{+0.73}_{-0.09}$ |
| HE0108-4743 | 17.79065 | −47.460342 | 0.024 | −5 | 5.7 | 8.09 | $9.77^{+0.18}_{-0.1}$ | 40.567 ± 0.002 | 37.67 ± 0.05 | ... | $0.52^{+0.02}_{-0.02}$ |
| NGC424 | 17.8652 | −38.0835 | 0.011 | 0.1 | 7.49 | 8.5 | $10.49^{+0.3}_{-0.3}$ | 40.815 ± 0 | ... | 38.27 ± 0.01 | $-0.22^{+0.08}_{-0.08}$ |
| SDSS J011221.82+150039.0 | 18.09094775 | 15.01090053 | 0.029 | −5 | 6.29 | 8.64 | $10.19^{+0.41}_{-0.49}$ | 38.717 ± 0.121 | 38.81 ± 0.02 | ... | $-0.89^{+0.63}_{-0.14}$ |
| Mrk975 | 18.4627 | 13.2718 | 0.05 | 2.7 | 7.45 | 10.14 | $10.98^{+0.3}_{-0.3}$ | 41.507 ± 0 | 38.84 ± 0.02 | 38.99 ± 0.02 | $1.11^{+0.04}_{-0.04}$ |
| IC1657 | 18.5292 | −32.6509 | 0.012 | 3.8 | 7.68 | 9.39 | $10.62^{+0.3}_{-0.3}$ | 39.769 ± 0.012 | 38.38 ± 0.02 | ... | $0.5^{+0.05}_{-0.05}$ |
| NGC454E | 18.6039 | −55.397 | 0.012 | −1 | 7.29 | 8.61 | $10.45^{+0.3}_{-0.3}$ | 40.205 ± 0.009 | ... | ... | $0.13^{+0}_{0}$ |
| SDSS J011501.75+152448.6 | 18.757321 | 15.41351941 | 0.031 | 1 | 7.19 | 8.4 | $10.33^{+0.4}_{-0.48}$ | 39.287 ± 0.047 | ... | ... | $-1.06^{+0.7}_{-0.1}$ |
| HE0114-0015 | 19.264947 | 0.007614 | 0.046 | 0.5 | 6.45 | 8.29 | $10.47^{+0.13}_{-0.19}$ | 40.479 ± 0 | 38.04 ± 0.02 | ... | $0.5^{+0.03}_{-0.03}$ |
| HE0119-0118 | 20.499281 | −1.040023 | 0.055 | 2 | 7.2 | 8.89 | $10.91^{+0.02}_{-0.06}$ | 41.728 ± 0.001 | ... | 38.79 ± 0.07 | $1.37^{+0.01}_{-0.02}$ |
| NGC526A | 20.9765 | −35.0654 | 0.019 | −2 | 8.06 | 8.63 | $10.48^{+0.3}_{-0.3}$ | 40.946 ± 0.002 | 39.16 ± 0.03 | 38.15 ± 0.02 | $-0.49^{+0.08}_{-0.08}$ |
| NGC513 | 21.1117 | 33.7995 | 0.019 | 5.5 | 7.63 | 9.61 | $10.86^{+0.3}_{-0.3}$ | 40.632 ± 0 | 38.57 ± 0.02 | ... | $0.86^{+0.3}_{-0.3}$ |
| Mrk359 | 21.8855 | 19.1788 | 0.017 | −2 | 6.05 | 9.18 | $10.5^{+0.3}_{-0.3}$ | 40.495 ± 0.002 | ... | 37.53 ± 0.11 | $0.37^{+0.04}_{-0.04}$ |
| MCG-3-4-72 | 22.028 | −18.8086 | 0.043 | −1 | 7.03 | 9.32 | $10.74^{+0.3}_{-0.3}$ | 41.813 ± 0 | 37.43 ± 0.09 | 38.38 ± 0.09 | $0.08^{+0.04}_{-0.04}$ |
| Z459-58 | 22.1018 | 16.4593 | 0.038 | 0.1 | 8.1 | 9.77 | $10.83^{+0.3}_{-0.3}$ | 41.017 ± 0.001 | 38.48 ± 0.02 | 38.72 ± 0.04 | $0.99^{+0.01}_{-0.01}$ |
| ESO244-30 | 22.4632 | −42.3265 | 0.025 | 4 | 7.08 | 9.61 | $10.66^{+0.3}_{-0.3}$ | 39.75 ± 0.045 | 39 ± 0.03 | ... | $0.81^{+0.3}_{-0.3}$ |
| ESO353-9 | 22.96 | −33.1193 | 0.016 | 3.7 | 7.96 | 9.64 | $10.68^{+0.3}_{-0.3}$ | 40.156 ± 0.007 | ... | 38.35 ± 0.01 | $0.72^{+0.03}_{-0.03}$ |
| SDSS J014042.68+133304.6 | 25.17782034 | 13.5512926 | 0.045 | −1.5 | 8.38 | 8.69 | $10.87^{+0.39}_{-0.52}$ | 39.339 ± 0.14 | ... | ... | $-0.98^{+0.76}_{-0.08}$ |
| CSRG165 | 27.3427 | −50.2521 | 0.03 | 0 | 7.55 | 8.59 | $10.81^{+0.3}_{-0.3}$ | 40.468 ± 0.027 | ... | ... | $0.09^{+0.05}_{-0.05}$ |
| ESO354-4 | 27.9244 | −36.1878 | 0.033 | 2.6 | 7.89 | 9.41 | $10.94^{+0.3}_{-0.3}$ | 41.127 ± 0.002 | ... | ... | $0.47^{+0.3}_{-0.3}$ |
| SDSS J015651.99+131246.0 | 29.21663506 | 13.21277794 | 0.044 | 1.1 | 8.62 | 8.69 | $10.9^{+0.39}_{-0.52}$ | 39.703 ± 0.053 | 37.76 ± 0.04 | ... | $-1.25^{+0.75}_{-0.09}$ |
| SDSS J015707.32+145543.5 | 29.28047659 | 14.92876888 | 0.043 | −3 | 7.33 | 8.51 | $10.87^{+0.4}_{-0.47}$ | 39.604 ± 0.065 | ... | ... | $-1.18^{+0.77}_{-0.07}$ |
| SDSS J015709.80+143259.5 | 29.29079878 | 14.54988781 | 0.026 | 1 | 7.73 | 8.81 | $11.01^{+0.39}_{-0.52}$ | 39.319 ± 0.046 | ... | ... | $-1.12^{+0.78}_{-0.09}$ |
| SDSS J015816.23+141747.9 | 29.56764059 | 14.29662687 | 0.026 | −4 | 8.15 | 8.51 | $10.8^{+0.38}_{-0.51}$ | 39.382 ± 0.046 | 36.93 ± 0.17 | ... | $-1.37^{+0.74}_{-0.09}$ |
| PG0157+001 | 29.959214 | 0.394615 | 0.164 | −5 | 8.31 | 10.37 | $11.53^{+0.3}_{-0.3}$ | 43.08 ± 0.016 | ... | 40.33 ± 0.01 | $2.33^{+0.03}_{-0.05}$ |

**Table 1**
(Continued)

| Name (1) | R.A. (2) | Decl. (3) | $z_*$ (4) | T-type (5) | $\log M_{BH}/M_\odot$ (6) | $\log M_{H_2}/M_\odot$ (7) | $\log M_*/M_\odot$ (8) | $\log L_{[O\ III]}$ (9) | $\log L_{[1.4\ GHz]}$ (10) | $\log L_{[3\ GHz]}$ (11) | log SFR (12) |
|---|---|---|---|---|---|---|---|---|---|---|---|
| UGC1479 | 30.0794 | 24.4738 | 0.017 | 3.5 | 7.51 | 9.32 | $10.6^{+0.3}_{-0.3}$ | 40.391 ± 0.001 | ... | 38.07 ± 0.03 | $0.46^{+0.08}_{-0.08}$ |
| NGC788 | 30.2769 | −6.8159 | 0.014 | 0 | 8.18 | 8.66 | $10.89^{+0.3}_{-0.3}$ | 40.727 ± 0 | 38.68 ± 0 | 37.49 ± 0.04 | $-0.12^{+0.3}_{-0.3}$ |
| SDSS J020939.47+135859.4 | 32.41449658 | 13.98318085 | 0.049 | 2 | 7.56 | 9.57 | $11.33^{+0.39}_{-0.49}$ | 39.32 ± 0.091 | ... | ... | $-0.9^{+0.81}_{-0.07}$ |
| Fairall377 | 32.7189 | −49.6985 | 0.048 | 5 | 8.27 | 10.05 | $11.04^{+0.3}_{-0.3}$ | 40.922 ± 0.024 | ... | ... | $1.11^{+0.3}_{-0.3}$ |
| SDSS J021130.77+141801.9 | 32.87824485 | 14.30051687 | 0.027 | −0.1 | 7.78 | 8.39 | $10.77^{+0.38}_{-0.5}$ | 39.279 ± 0.047 | ... | ... | $-1.31^{+0.75}_{-0.09}$ |
| SDSS J021131.43+141202.4 | 32.88100178 | 14.20064172 | 0.025 | 0 | 6.98 | 8.26 | $10.33^{+0.39}_{-0.49}$ | 39.151 ± 0.043 | 38.25 ± 0.05 | ... | $-1.48^{+0.77}_{-0.08}$ |
| SDSS J021337.50+134341.2 | 33.40626002 | 13.72812133 | 0.041 | 1 | 8.31 | 8.83 | $11.18^{+0.38}_{-0.53}$ | 39.515 ± 0.054 | 37.36 ± 0.07 | ... | $-1.31^{+0.75}_{-0.09}$ |

**Note.** Column (1): name. Column (2): the R.A. in decimal degrees. Column (3): (delineation) in decimal degrees. Column (4): redshift. Column (5): morphological parameters. Column (6): logarithm of black mass, the $1\sigma$ uncertainty is 0.4 dex. Column (7): logarithm of molecular gas mass inferred from CO measurements; the $1\sigma$ uncertainty is 0.3 dex (J. Shangguan et al. 2020b; J. Molina et al. 2023a). Column (8): logarithm of stellar mass; the $1\sigma$ uncertainties of stellar mass for BAT-AGN and PG quasars are 0.3 dex (J. Shangguan et al. 2018; J. Molina et al. 2023a). The uncertainty in the stellar masses of the remaining samples comes from references to corresponding stellar masses. Column (9): the [O III] luminosity in units erg $s^{-1}$. Column (10): the 1.4 GHz radio luminosity in units erg $s^{-1}$. Column (11): the 3 GHz radio luminosity in units erg $s^{-1}$. Column (12): logarithm of star formation rates and error come from the corresponding reference. This table is available in its entirety in machine-readable form.

(This table is available in its entirety in machine-readable form in the online article.)

**Table 2**
Best-fitted Linear Relations between Luminosity and Physical Parameters for AGNs

| | Type | $\alpha$ | $\beta$ | $r$ | $P$ | Intrinsic Scatter |
|---|---|---|---|---|---|---|
| $\log L_{1.4\ \mathrm{GHz}} - \log M_{\mathrm{H_2}}$ | all | $0.94 \pm 0.10$ | $29.83 \pm 0.96$ | 0.55 | $4.43 \times 10^{-20}$ | 0.25 |
| | Early | $1.00 \pm 0.13$ | $29.39 \pm 1.24$ | 0.60 | $2.3 \times 10^{-11}$ | 0.32 |
| | Late | $1.23 \pm 0.35$ | $26.92 \pm 3.31$ | 0.57 | $6.86 \times 10^{-13}$ | 0.13 |
| $\log L_{3\ \mathrm{GHz}} - \log M_{\mathrm{H_2}}$ | all | $1.18 \pm 0.14$ | $27.51 \pm 1.29$ | 0.59 | $3.6 \times 10^{-18}$ | 0.21 |
| | Early | $1.37 \pm 0.15$ | $25.81 \pm 1.41$ | 0.75 | $7.8 \times 10^{-17}$ | 0.15 |
| | Late | $0.90 \pm 0.28$ | $30.02 \pm 2.71$ | 0.42 | $2.19 \times 10^{-5}$ | 0.24 |
| $\log L_{[\mathrm{O\ III}]} - \log M_{\mathrm{H_2}}$ | all | $0.96 \pm 0.12$ | $31.64 \pm 1.12$ | 0.36 | $9.28 \times 10^{-14}$ | 0.78 |
| | Early | $1.21 \pm 0.14$ | $29.75 \pm 1.34$ | 0.53 | $8.3 \times 10^{-16}$ | 0.72 |
| | Late | $1.03 \pm 0.23$ | $30.66 \pm 2.15$ | 0.33 | $1.02 \times 10^{-6}$ | 0.72 |
| $\log L_{5100\ \text{Å}} - \log M_{\mathrm{H_2}}$ | all | $1.23 \pm 0.13$ | $29.78 \pm 1.20$ | 0.45 | $7.4 \times 10^{-20}$ | 0.38 |
| | Early | $1.59 \pm 0.21$ | $26.73 \pm 1.91$ | 0.59 | $5.6 \times 10^{-17}$ | 0.46 |
| | Late | $1.18 \pm 0.21$ | $30.00 \pm 1.94$ | 0.40 | $1.8 \times 10^{-9}$ | 0.34 |
| $\log L_{1.4\ \mathrm{GHz}} - \log M_{\mathrm{H_2}}/M_*$ | all | $1.27 \pm 0.22$ | $40.32 \pm 0.29$ | 0.35 | $3.6 \times 10^{-8}$ | 0.25 |
| | Early | $1.11 \pm 0.19$ | $40.33 \pm 0.29$ | 0.43 | $5.22 \times 10^{-6}$ | 0.34 |
| | Late | $4.30 \pm 1.56$ | $43.91 \pm 1.97$ | 0.31 | $2.72 \times 10^{-4}$ | 0.06 |
| $\log L_{3\ \mathrm{GHz}} - \log M_{\mathrm{H_2}}/M_*$ | all | $1.74 \pm 0.30$ | $40.82 \pm 0.38$ | 0.41 | $1.06 \times 10^{-8}$ | 0.18 |
| | Early | $1.63 \pm 0.29$ | $40.85 \pm 0.41$ | 0.54 | $1.02 \times 10^{-7}$ | 0.22 |
| | Late | $2.52 \pm 1.61$ | $41.63 \pm 2.00$ | 0.28 | 0.005 | 0.16 |
| $\log L_{[\mathrm{O\ III}]} - \log M_{\mathrm{H_2}}/M_*$ | all | $1.14 \pm 0.21$ | $42.14 \pm 0.31$ | 0.30 | $5.9 \times 10^{-10}$ | 0.82 |
| | Early | $1.37 \pm 0.19$ | $42.88 \pm 0.32$ | 0.53 | $2.5 \times 10^{-15}$ | 0.63 |
| | Late | $0.48 \pm 0.14$ | $40.97 \pm 0.21$ | 0.21 | $2.15 \times 10^{-3}$ | 0.82 |
| $\log L_{5100\ \text{Å}} - \log M_{\mathrm{H_2}}/M_*$ | all | $2.72 \pm 0.38$ | $45.01 \pm 0.54$ | 0.44 | $1.11 \times 10^{-19}$ | 0.15 |
| | Early | $1.91 \pm 0.24$ | $44.13 \pm 0.36$ | 0.60 | $1.14 \times 10^{-17}$ | 0.17 |
| | Late | $4.49 \pm 1.53$ | $47.16 \pm 2.08$ | 0.35 | $2.2 \times 10^{-7}$ | 0.16 |
| $\log L_{1.4\ \mathrm{GHz}} - \log M_{\mathrm{BH}}/M_\odot$ | all | $0.23 \pm 0.06$ | $36.89 \pm 0.47$ | 0.28 | $5.1 \times 10^{-6}$ | 0.39 |
| | Early | $0.36 \pm 0.12$ | $35.99 \pm 0.90$ | 0.35 | $2.5 \times 10^{-4}$ | 0.58 |
| | Late | $0.12 \pm 0.07$ | $37.69 \pm 0.53$ | 0.23 | 0.008 | 0.25 |
| $\log L_{3\ \mathrm{GHz}} - \log M_{\mathrm{BH}}/M_\odot$ | all | $0.16 \pm 0.08$ | $37.39 \pm 0.60$ | 0.17 | 0.02 | 0.45 |
| | Early | $0.35 \pm 0.13$ | $35.90 \pm 1.03$ | 0.32 | $2.8 \times 10^{-3}$ | 0.58 |
| | Late | $-0.01 \pm 0.09$ | $38.63 \pm 0.72$ | 0.011 | 0.48 | 0.33 |
| $\log L_{[\mathrm{O\ III}]} - \log M_{\mathrm{BH}}/M_\odot$ | all | $0.22 \pm 0.07$ | $38.85 \pm 0.56$ | 0.18 | $3.2 \times 10^{-4}$ | 0.95 |
| | Early | $0.19 \pm 0.12$ | $39.23 \pm 0.92$ | 0.11 | 0.13 | 0.19 |
| | Late | $0.17 \pm 0.09$ | $39.07 \pm 0.71$ | 0.21 | $1.6 \times 10^{-3}$ | 0.85 |
| $\log L_{5100\ \text{Å}} - \log M_{\mathrm{BH}}/M_\odot$ | all | -0.14 $\pm$ 0.06 | $42.15 \pm 0.46$ | -0.06 | 0.186 | 0.59 |
| | Early | $-0.17 \pm 0.12$ | $42.49 \pm 0.90$ | $-0.07$ | 0.35 | 0.75 |
| | Late | $-0.15 \pm 0.07$ | $42.16 \pm 0.54$ | -0.08 | 0.24 | 0.48 |
| $\log L_{1.4\ \mathrm{GHz}} - \log M_*/M_\odot$ | all | $0.67 \pm 0.19$ | $31.48 \pm 2.09$ | 0.36 | $1.2 \times 10^{-8}$ | 0.39 |
| | Early | $0.63 \pm 0.33$ | $31.99 \pm 3.61$ | 0.29 | $2.6 \times 10^{-3}$ | 0.62 |
| | Late | $0.64 \pm 0.21$ | $31.74 \pm 2.26$ | 0.42 | $3.6 \times 10^{-7}$ | 0.23 |
| $\log L_{3\ \mathrm{GHz}} - \log M_*/M_\odot$ | all | $0.55 \pm 0.26$ | $32.72 \pm 2.73$ | 0.32 | $9.2 \times 10^{-6}$ | 0.45 |
| | Early | $0.71 \pm 0.39$ | $31.11 \pm 4.21$ | 0.36 | $5.3 \times 10^{-4}$ | 0.61 |
| | Late | $0.41 \pm 0.31$ | $34.15 \pm 3.32$ | 0.26 | 0.009 | 0.32 |
| $\log L_{[\mathrm{O\ III}]} - \log M_*/M_\odot$ | all | $0.45 \pm 0.28$ | $35.72 \pm 2.97$ | 0.11 | 0.019 | 0.96 |
| | Early | $0.03 \pm 0.41$ | $40.33 \pm 4.43$ | 0.002 | 0.502 | 0.99 |
| | Late | $0.85 \pm 0.36$ | $31.21 \pm 3.86$ | 0.23 | $4.9 \times 10^{-4}$ | 0.82 |
| $\log L_{5100\ \text{Å}} - \log M_*/M_\odot$ | all | -0.17 $\pm$ 0.26 | $42.96 \pm 2.77$ | 0.018 | 0.47 | 0.61 |
| | Early | $-0.63 \pm 0.44$ | $47.90 \pm 4.71$ | 0.07 | 0.34 | 0.74 |
| | Late | $0.24 \pm 0.31$ | $38.46 \pm 3.26$ | 0.11 | 0.09 | 0.49 |

| | Type | $\alpha$ | $\beta$ | r | P | Intrinsic Scatter |
|---|---|---|---|---|---|---|
| $\log L_{1.4\ \mathrm{GHz}} - \log\ \mathrm{SFR}$ | all | $0.82 \pm 0.06$ | $38.13 \pm 0.05$ | 0.63 | $2.92 \times 10^{-28}$ | 0.22 |
| | Early | $0.86 \pm 0.08$ | $38.24 \pm 0.07$ | 0.69 | $3.11 \times 10^{-16}$ | 0.29 |
| | Late | $0.77 \pm 0.09$ | $38.08 \pm 0.07$ | 0.60 | $1.02 \times 10^{-14}$ | 0.16 |
| $\log L_{3\ \mathrm{GHz}} - \log\ \mathrm{SFR}$ | all | $0.76 \pm 0.06$ | $38.14 \pm 0.05$ | 0.65 | $4.21 \times 10^{-23}$ | 0.23 |
| | Early | $0.81 \pm 0.08$ | $38.25 \pm 0.07$ | 0.72 | $1.41 \times 10^{-14}$ | 0.24 |
| | Late | $0.72 \pm 0.11$ | $38.06 \pm 0.09$ | 0.56 | $2.68 \times 10^{-9}$ | 0.69 |
| $\log L_{[\mathrm{O\ III}]} - \log\ \mathrm{SFR}$ | all | $0.68 \pm 0.05$ | $40.33 \pm 0.05$ | 0.48 | $1.57 \times 10^{-24}$ | 0.70 |
| | Early | $0.83 \pm 0.06$ | $40.63 \pm 0.05$ | 0.68 | $2.31 \times 10^{-27}$ | 0.52 |
| | Late | $0.66 \pm 0.09$ | $40.04 \pm 0.07$ | 0.40 | $1.08 \times 10^{-9}$ | 0.55 |

**Table 2**
(Continued)

| | Type | $\alpha$ | $\beta$ | $r$ | $P$ | Intrinsic Scatter |
|---|---|---|---|---|---|---|
| $\log L_{5100\,Å} - \log$ SFR | all | $0.64 \pm 0.03$ | $40.97 \pm 0.04$ | 0.56 | $5.62 \times 10^{-32}$ | 0.38 |
| | Early | $0.77 \pm 0.06$ | $41.21 \pm 0.05$ | 0.70 | $5.94 \times 10^{-26}$ | 0.35 |
| | Late | $0.64 \pm 0.07$ | $40.78 \pm 0.05$ | 0.50 | $1.74 \times 10^{-14}$ | 0.33 |

**Note.** $\log Y = \alpha \log X + \beta$. The all is the both early-type and late-type galaxies. Early is the early-type galaxies ($T \leqslant 0$). Late is the late-type galaxies ($T > 0$). $r$ and $P$ are Spearman correlation coefficient and probability for the null hypothesis of no correlation.

### 2.2. The Radio Luminosity

We obtain 1.4 GHz radio flux from the NASA/IPAC Extragalactic Database (J. J. Condon et al. 1998; or Faint Images of the Radio Sky at Twenty cm; R. L. White et al. 1997) and 3 GHz radio flux from the Very Large Array Sky Survey catalogs (Y. A. Gordon et al. 2021) for our sample, respectively. The radio luminosity was estimated by using the formula $L_\nu = 4\pi d_L^2 S_\nu$, where the luminosity distance $d_L(z)$ is calculated as $d_L(z) = \frac{c}{H_0}(1 + z) \int_0^z [\Omega_\Lambda + \Omega_m(1 + z')^3]^{-1/2} dz'$ (T. M. Venters et al. 2009). We make a K-correction for the radio flux using $S_\nu = S_\nu^{\rm obs}(1 + z)^{\alpha - 1}$ and $\alpha = 0$ (P. Cassaro et al. 1999; A. A. Abdo et al. 2010). The low-resolution radio emission may come from contributions from star formation (L. C. Ho 1999; L. C. Ho & C. Y. Peng 2001; L. C. Ho & J. S. Ulvestad 2001; J. S. Ulvestad & L. C. Ho 2001; L. C. Ho 2002; D. V. Lal & L. C. Ho 2010). To avoid radio emission from star formation, we only choose the radio excess sources. In AGNs with $\log L_1.4{\rm GHz}/L_1.4{\rm GHz,\,SF}$ $(\log L_{3\,\rm GHz}/L_{3\,\rm GHz,SF})$ significantly larger than zero, the detected radio emission cannot be explained by star formation, and thus likely arises from nuclear activity (e.g., jets). The radio excess sources are defined by using $\log L_{1.4\,\rm GHz}/L_{1.4\,\rm GHz,\,SF}$ $(\log L_{3\,\rm GHz}/L_{3\,\rm GHz,SF}) > 0$ (E. J. Murphy et al. 2011; M. Bonzini et al. 2015; V. Smolčić et al. 2017; J. S. Elford et al. 2024). The $L_{1.4\,\rm GHz}$, SF ($L_{3\,\rm GHz,SF}$) is estimated by using the following formula

$$\left(\frac{L_\nu}{\rm erg\ s^{-1}Hz^{-1}}\right) = 10^{27}\left(\frac{{\rm SFR}_\nu}{M_\odot\,{\rm yr}^{-1}}\right)\left[2.18\left(\frac{T_e}{10^4 K}\right)^{0.45} \left(\frac{\nu}{\rm GHz}\right)^{-0.1} + 15.1\left(\frac{\nu}{\rm GHz}\right)^{\alpha^{\rm NT}}\right], \quad (2)$$

where $\nu$ is the observed frequency, and $T_e$ is the electron temperature. The $\alpha^{\rm NT}$ is the nonthermal spectral index. The $T_e = 10^4$ K and $\alpha^{\rm NT} = -0.8$ are adopted (E. J. Murphy et al. 2011; J. S. Elford et al. 2024).

### 2.3. The [O III] Luminosity and 5100 Å Luminosity

The [O III] luminosity of BAT-AGN is from the work of K. Oh et al. (2022). For the Palomar–Green quasars, [O III] luminosities have been measured in various studies (A. Alonso-Herrero et al. 1997; Y. Matsuoka 2012; A. Pennell et al. 2017; H.-Y. Liu et al. 2019; K. Oh et al. 2022). The [O III] luminosities of the CARS sample are provided by B. Husemann et al. (2022). Additionally, the QFS samples include measured [O III] luminosities as reported by J. R. Mullaney et al. (2013). For the xGASS and MaNGA samples, the [O III] luminosities are obtained from the NASA-Sloan Atlas catalog. J. E. Greene & L. C. Ho (2005) discovered a tight connection between 5100 Å luminosity and H$\beta$ luminosity. B. Husemann et al. (2022) used the H$\beta$ luminosity to estimate the 5100 Å luminosity. Following them, we also use the H$\beta$ line luminosity to estimate the 5100 Å luminosity for our sample.

### 2.4. The Morphology

The morphology parameter T-type is obtained from the HyperLEDA database (D. Makarov et al. 2014). This parameter can take noninteger values, as it often represents the average of multiple estimates found in the literature for most objects. Galaxies classified as early-type galaxies have a T-type value of $T \leqslant 0$, while late-type galaxies are characterized by $T > 0$. The mass of the bulge and the corresponding error for the AGNs is from the work of T. Bohn et al. (2020), W.-H. Bian & Y.-H. Zhao (2003), R. Läsker et al. (2016), and J. T. Mendel et al. (2014).

## 3. Result and Discussion

### 3.1. Relation between Molecular Gas Mass and AGN Power

The origin of the activity of supermassive black holes and their impact on host galaxies remains a highly intriguing topic in extragalactic astronomy. The detection of cold molecular gas in massive galaxies may suggest that this gas serves as fuel for their AGNs. It would be difficult to fuel strong AGN activity when a galaxy has no cold gas, although weak AGN activity, in the form of radiatively inefficient low-ionization nuclear emission-line regions, can still be powered by small amounts of gas, including hot gas (L. C. Ho et al. 2003; L. C. Ho 2008, 2009; T. Izumi et al. 2016). At present, it is unclear whether the mechanism that fuels the AGNs in early-type galaxies is similar to that in late-type galaxies. To test these hypotheses, we study the relation between molecular gas mass and AGN power. We conduct a linear regression analysis using LINMIX (B. C. Kelly 2007), a method that accommodates measurement uncertainties in both variables and simultaneously accounts for intrinsic, random scatter. We use PYMCCORRELATION (G. C. Privon et al. 2020) with bootstrapping iterations to obtain Spearman's rank correlation coefficient and the corresponding $p$-value (see Table 2). Many authors also use the PYMCCORRELATION to obtain the correlation coefficients and $p$-value between the two variables (e.g., V. U et al. 2022; M. Bierschenk et al. 2024). We also use the Python package pingouin.partial_corr[11] to do a partial correlation analysis and get Spearman's rank correlation coefficient. We assume $p < 0.05$ to determine if the quantities are correlated.

We define the molecular gas content as the ratio of molecular gas mass to stellar mass, denoted as $M_{\rm H_2}/M_*$. Initially, we examine the relationships between radio luminosity and [O III] luminosity and molecular gas mass in Figure 1. We employ a linear regression analysis to investigate the relationship between radio luminosity and [O III] luminosity and molecular gas mass for AGNs, respectively. A significant correlation is observed between radio luminosity and [O III] luminosity and molecular gas mass for AGNs (Spearman correlation coefficient $r_{1.4\ \mathrm{GHz}} = 0.55$, $P = 4.43 \times 10^{-20}$;

[11] https://en.wikipedia.org/wiki/Partial_correlation

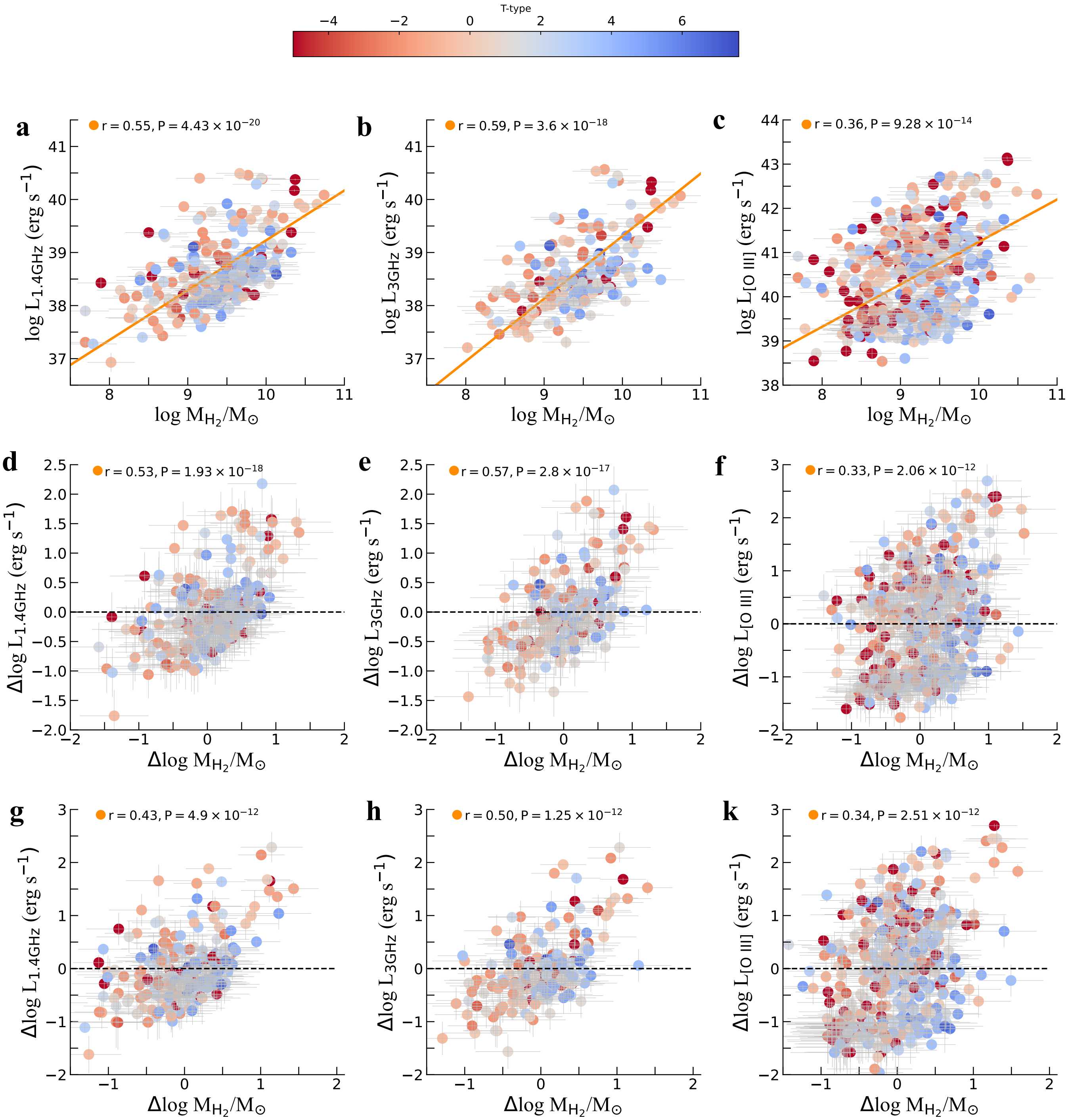

**Figure 1.** Relation between AGN luminosity and cold molecular gas mass of AGNs. (a), (b) ,(c): $L_{1.4\ \mathrm{GHz}}$–$M_{\mathrm{H_2}}$ (a), $L_{3\ \mathrm{GHz}}$–$M_{\mathrm{H_2}}$ (b), and $L_{[\mathrm{O\ III}]}$–$M_{\mathrm{H_2}}$ (c) correlations. The color bar indicates morphological T types of host galaxies of AGNs (smaller values being more early-type and larger values more late-type galaxies). The orange line is the best-fitted linear relation, taking into account the uncertainties of both variables. (d), (e), (f), (g), (h), (k): comparison of the partial correlation of $L_{1.4\ \mathrm{GHz}}$–$M_{\mathrm{H_2}}$ (while controlling for $M_{\mathrm{BH}}$) (d), $L_{3\ \mathrm{GHz}}$–$M_{\mathrm{H_2}}$ (while controlling for $M_{\mathrm{BH}}$) (e), $L_{[\mathrm{O\ III}]}$–$M_{\mathrm{H_2}}$ (while controlling for $M_{\mathrm{BH}}$) (f), $L_{1.4\ \mathrm{GHz}}$–$M_{\mathrm{H_2}}$ (while controlling for $M_*$) (g), $L_{3\ \mathrm{GHz}}$–$M_{\mathrm{H_2}}$ (while controlling for $M_*$) (h), and $L_{[\mathrm{O\ III}]}$–$M_{\mathrm{H_2}}$ (while controlling for $M_*$) (k). The $x$- and $y$-axes are the residual in $M_{\mathrm{H_2}}$, $L_{1.4\ \mathrm{GHz}}$, $L_{3\ \mathrm{GHz}}$, and $L_{[\mathrm{O\ III}]}$ after removing their dependence on $M_{\mathrm{BH}}$ in the middle panel, and after removing their dependence on $M_*$ in the bottom panel: $\Delta\log L_{1.4\,\mathrm{GHz}} = \log L_{1.4\,\mathrm{GHz}} - \log L_{1.4\,\mathrm{GHz}}(M_{\mathrm{BH}})$, $\Delta\log L_{3\,\mathrm{GHz}} = \log L_{3\,\mathrm{GHz}} - \log L_{3\,\mathrm{GHz}}(M_{\mathrm{BH}})$, $\Delta\log L_{[\mathrm{O\,III}]} = \log L_{[\mathrm{O\,III}]} - \log L_{[\mathrm{O\,III}]}(M_{\mathrm{BH}})$, and $\Delta\log M_{\mathrm{H_2}} = \log M_{\mathrm{H_2}} - \log M_{\mathrm{H_2}}(M_{\mathrm{BH}})$ in the middle panel, and $\Delta\log L_{1.4\,\mathrm{GHz}} = \log L_{1.4\,\mathrm{GHz}} - \log L_{1.4\,\mathrm{GHz}}(M_*)$, $\Delta\log L_{3\,\mathrm{GHz}} = \log L_{3\,\mathrm{GHz}} - \log L_{3\,\mathrm{GHz}}(M_*)$, $\Delta\log L_{[\mathrm{O\,III}]} = \log L_{[\mathrm{O\,III}]} - \log L_{[\mathrm{O\,III}]}(M_*)$, and $\Delta\log M_{\mathrm{H_2}} = \log M_{\mathrm{H_2}} - \log M_{\mathrm{H_2}}(M_*)$ in the bottom panel. The dashed line shows zero correlation, that is, there is no intrinsic correlation between the two quantities. The Spearman correlation coefficients between the two corresponding variables are shown in each panel. The error bars refer to $1\sigma$ measurement errors.

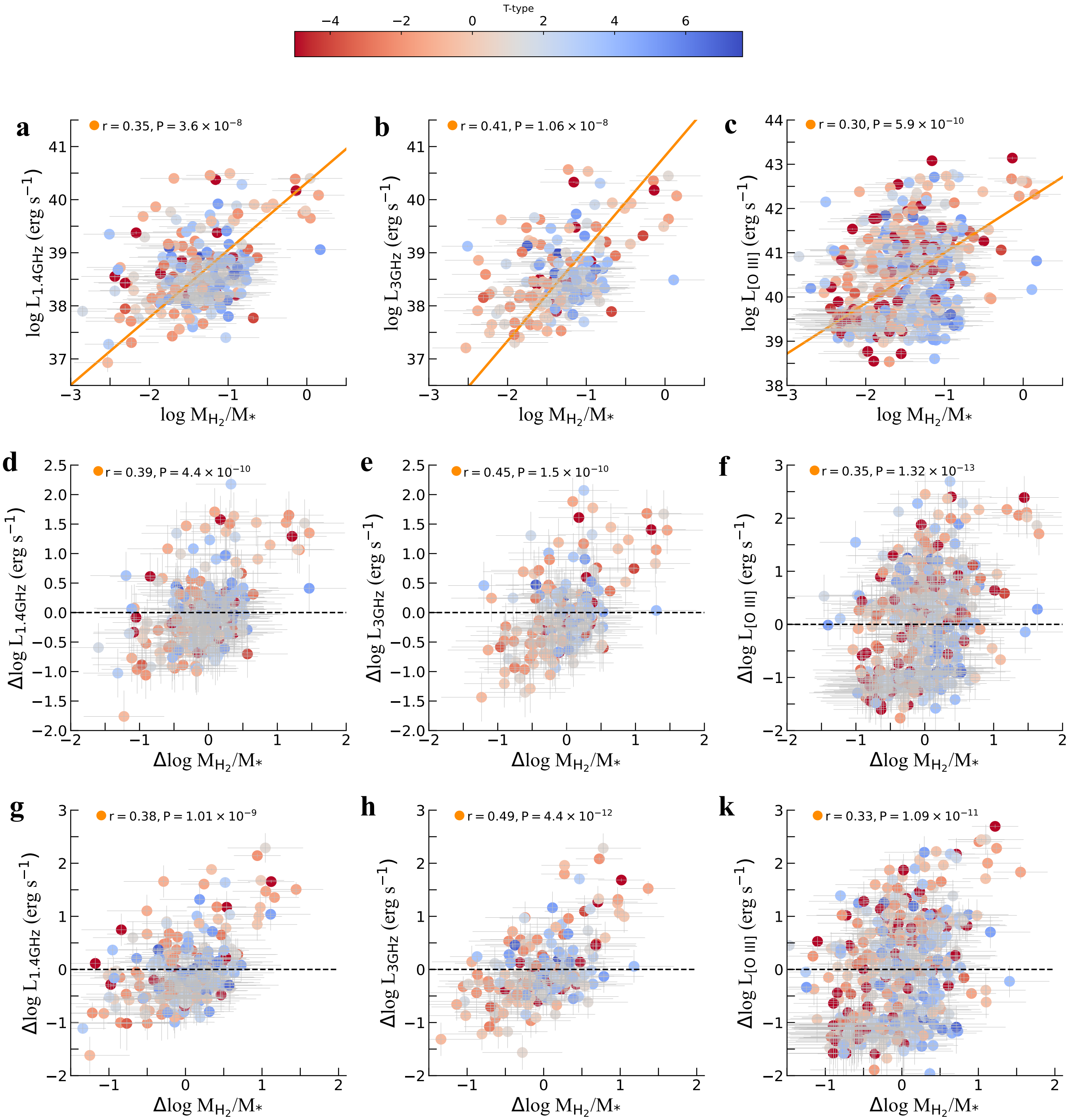

**Figure 2.** Relation between AGN luminosity and molecular gas content for AGNs. (a), (b), (c): $L_{1.4\ \mathrm{GHz}}$–$M_{\mathrm{H_2}}/M_*$ (a), $L_{3\ \mathrm{GHz}}$–$M_{\mathrm{H_2}}/M_*$ (b), and $L_{[\mathrm{O\ III}]}$–$M_{\mathrm{H_2}}/M_*$ (c) correlations. The color bar indicates morphological T types of host galaxies of AGNs (smaller values being more early-type and larger values more late-type galaxies). The orange lines are the best-fitted linear relation, taking into account the uncertainties of both variables. (d), (e), (f), (g), (h), (k): comparison of the partial correlation of $L_{1.4\ \mathrm{GHz}}$–$M_{\mathrm{H_2}}/M_*$ (while controlling for $M_{\mathrm{BH}}$) (d), $L_{3\ \mathrm{GHz}}$–$M_{\mathrm{H_2}}/M_*$ (while controlling for $M_{\mathrm{BH}}$) (e), $L_{[\mathrm{O\ III}]}$–$M_{\mathrm{H_2}}/M_*$ (while controlling for $M_{\mathrm{BH}}$) (f), $L_{1.4\ \mathrm{GHz}}$–$M_{\mathrm{H_2}}/M_*$ (while controlling for $M_*$) (g), $L_{3\ \mathrm{GHz}}$–$M_{\mathrm{H_2}}/M_*$ (while controlling for $M_*$) (h), and $L_{[\mathrm{O\ III}]}$–$M_{\mathrm{H_2}}/M_*$ (while controlling for $M_*$) (k). The $x$- and $y$-axes are the residual in $M_{\mathrm{H_2}}/M_*$, $L_{1.4\ \mathrm{GHz}}$, $L_{3\ \mathrm{GHz}}$, and $L_{[\mathrm{O\ III}]}$ after removing their dependence on $M_{\mathrm{BH}}$ in the middle panel, and after removing their dependence on $M_*$ in the bottom panel: $\Delta \log Y = \log Y - \log Y(M_{\mathrm{BH}})$ and $\Delta \log M_{\mathrm{H_2}}/M_* = \log M_{\mathrm{H_2}}/M_* - \log M_{\mathrm{H_2}}/M_*(M_{\mathrm{BH}})$ in the middle panel, and $\Delta \log Y = \log Y - \log Y(M_*)$ and $\Delta \log M_{\mathrm{H_2}}/M_* = \log M_{\mathrm{H_2}}/M_* - \log M_{\mathrm{H_2}}/M_*(M_*)$ in the bottom panel. The dashed line shows zero correlation, that is, there is no intrinsic correlation between the two quantities. The Spearman correlation coefficients between the two corresponding variables are shown in each panel. The error bars refer to $1\sigma$ measurement errors.

$r_{3\ \mathrm{GHz}} = 0.59$, $P = 3.6 \times 10^{-18}$; $r_{[\mathrm{O\ III}]} = 0.36$, $P = 9.28 \times 10^{-14}$). More importantly, a partial correlation analysis controlling for black hole mass and stellar mass reveals that the relationship between radio luminosity and [O III] luminosity and molecular gas mass remains significant after removing dependencies on black hole mass ($r_{1.4\ \mathrm{GHz}} = 0.53$, $P = 1.93 \times 10^{-18}$; $r_{3\ \mathrm{GHz}} = 0.57$, $P = 2.8 \times 10^{-17}$; $r_{[\mathrm{O\ III}]} = 0.33$, $P = 2.06 \times 10^{-12}$; Figure 1, middle panels) and stellar mass ($r_{1.4\ \mathrm{GHz}} = 0.43$, $P = 4.9 \times 10^{-12}$; $r_{3\ \mathrm{GHz}} = 0.50$, $P = 1.25 \times 10^{-12}$; $r_{[\mathrm{O\ III}]} = 0.34$, $P = 2.51 \times 10^{-12}$; Figure 1, bottom panels). This indicates that radio luminosity and [O III] luminosity are still dependent on molecular gas mass. Additionally, a significant correlation exists between radio luminosity and [O III] luminosity and molecular gas content for AGNs ($r_{1.4\ \mathrm{GHz}} = 0.35$, $P = 3.6 \times 10^{-8}$; $r_{3\ \mathrm{GHz}} = 0.41$, $P = 1.06 \times 10^{-8}$; $r_{[\mathrm{O\ III}]} = 0.30$, $P = 5.9 \times 10^{-10}$). A partial correlation analysis controlling for black hole mass and stellar mass further confirms this dependence ($r_{1.4\ \mathrm{GHz}} = 0.39$, $P = 4.4 \times 10^{-10}$; $r_{3\ \mathrm{GHz}} = 0.45$, $P = 1.5 \times 10^{-10}$; $r_{[\mathrm{O\ III}]} = 0.35$, $P = 1.32 \times 10^{-13}$; Figure 2, middle panels) and stellar mass ($r_{1.4\ \mathrm{GHz}} = 0.38$, $P = 1.01 \times 10^{-9}$; $r_{3\ \mathrm{GHz}} = 0.49$, $P = 4.4 \times 10^{-12}$; $r_{[\mathrm{O\ III}]} = 0.33$, $P = 1.09 \times 10^{-11}$; Figure 2, bottom panels). R. D. Baldi et al. (2015) found that the slopes of the relation between 5 GHz radio luminosity and [O III]λ5007 line luminosity and molecular gas mass for 37 local early-type galaxies are 0.90 ± 0.40 and 0.95 ± 0.30, respectively. We found that the slopes of our samples were consistent with theirs within the margin of error (see Table 2). T. Izumi et al. (2016) found a correlation between molecular gas mass and the AGN activity for 10 Seyfert galaxies. J. Shangguan et al. (2020a) found that there is a correlation between the CO luminosity and the AGN luminosity and black hole mass for 40 Palomar-Green (PG) quasars. H. R. Russell et al. (2019) found a significant correlation between molecular gas mass and X-ray cavity jet power for 12 central cluster galaxies. M.-Y. Zhuang et al. (2021) also found a significant correlation between the molecular gas mass and AGN luminosity. Y. Fujita et al. (2024) also found a strong correlation between molecular gas mass and X-ray cavity jet power for 22 elliptical galaxies. Y. Chen et al. (2023) found a significant correlation between molecular gas fraction and CO luminosity and black hole spin for the early-type galaxies. These results suggest that the activity of AGNs depends on the amount of gas.

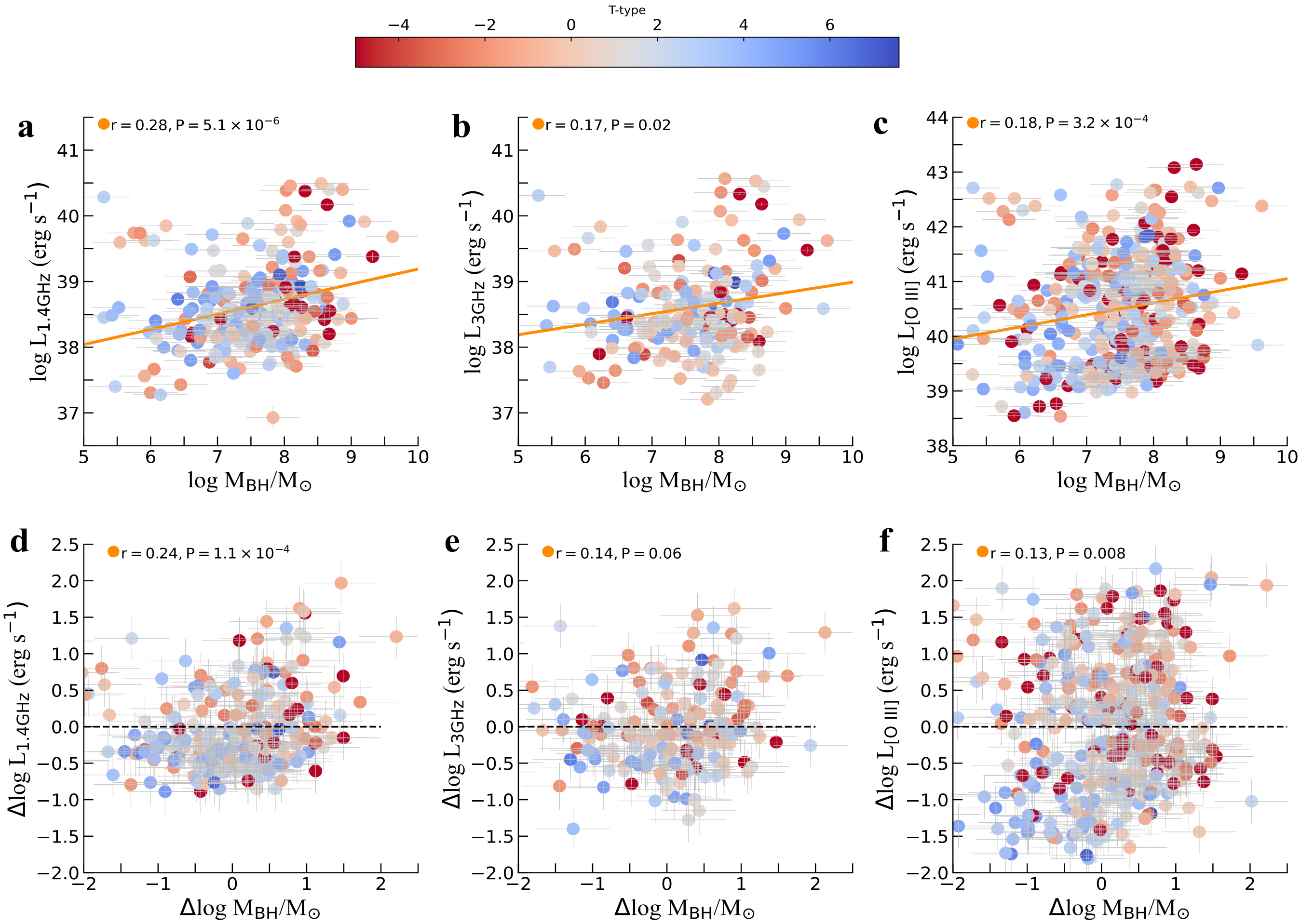

**Figure 3.** Relation between AGN luminosity and black hole mass of AGNs. (a), (b), (c): $L_{1.4\ \mathrm{GHz}}$–$M_{\mathrm{BH}}$ (a), $L_{3\ \mathrm{GHz}}$–$M_{\mathrm{BH}}$ (b), and $L_{[\mathrm{O\ III}]}$–$M_{\mathrm{BH}}$ (c) correlations. The color bar indicates morphological T types of host galaxies of AGNs (smaller values being more early-type and larger values more late-type galaxies). The orange lines are the best-fitted linear relation, taking into account the uncertainties of both variables. (d), (e), (f): comparison of the partial correlation of $L_{1.4\ \mathrm{GHz}}$–$M_{\mathrm{BH}}$ (while controlling for $M_{\mathrm{H_2}}$) (d), $L_{3\ \mathrm{GHz}}$–$M_{\mathrm{BH}}$ (while controlling for $M_{\mathrm{H_2}}$) (e), and $L_{[\mathrm{O\ III}]}$–$M_{\mathrm{BH}}$ (while controlling for $M_{\mathrm{H_2}}$) (f). The *x*- and *y*-axes are the residual in $M_{\mathrm{BH}}$, $L_{1.4\ \mathrm{GHz}}$, $L_{3\ \mathrm{GHz}}$, and $L_{[\mathrm{O\ III}]}$ after removing their dependence on $M_{\mathrm{H_2}}$ in the bottom panel: $\Delta \log Y = \log Y - \log Y(M_{\mathrm{H_2}})$ and $\Delta \log M_{\mathrm{BH}} = \log M_{\mathrm{BH}} - \log M_{\mathrm{BH}}(M_{\mathrm{H_2}})$.

### *3.2. Relation between Black Hole Mass and AGN Power*

Some studies suggest that the activity of AGNs may be related to the mass of black holes (L. C. Ho 2002, 2008; P. N. Best et al. 2005). However, some authors found that there is no correlation between the activity of AGNs and black hole mass (J.-H. Woo & C. M. Urry 2002; Y. Fujita et al. 2024). At the same time, there is a correlation between black hole mass and stellar mass, and bulge mass (A. Marconi & L. K. Hunt 2003; N. Häring & H.-W. Rix 2004; J. Kormendy & L. C. Ho 2013; T. M. Heckman & P. N. Best 2014; A. E. Reines & M. Volonteri 2015; M.-Y. Zhuang &

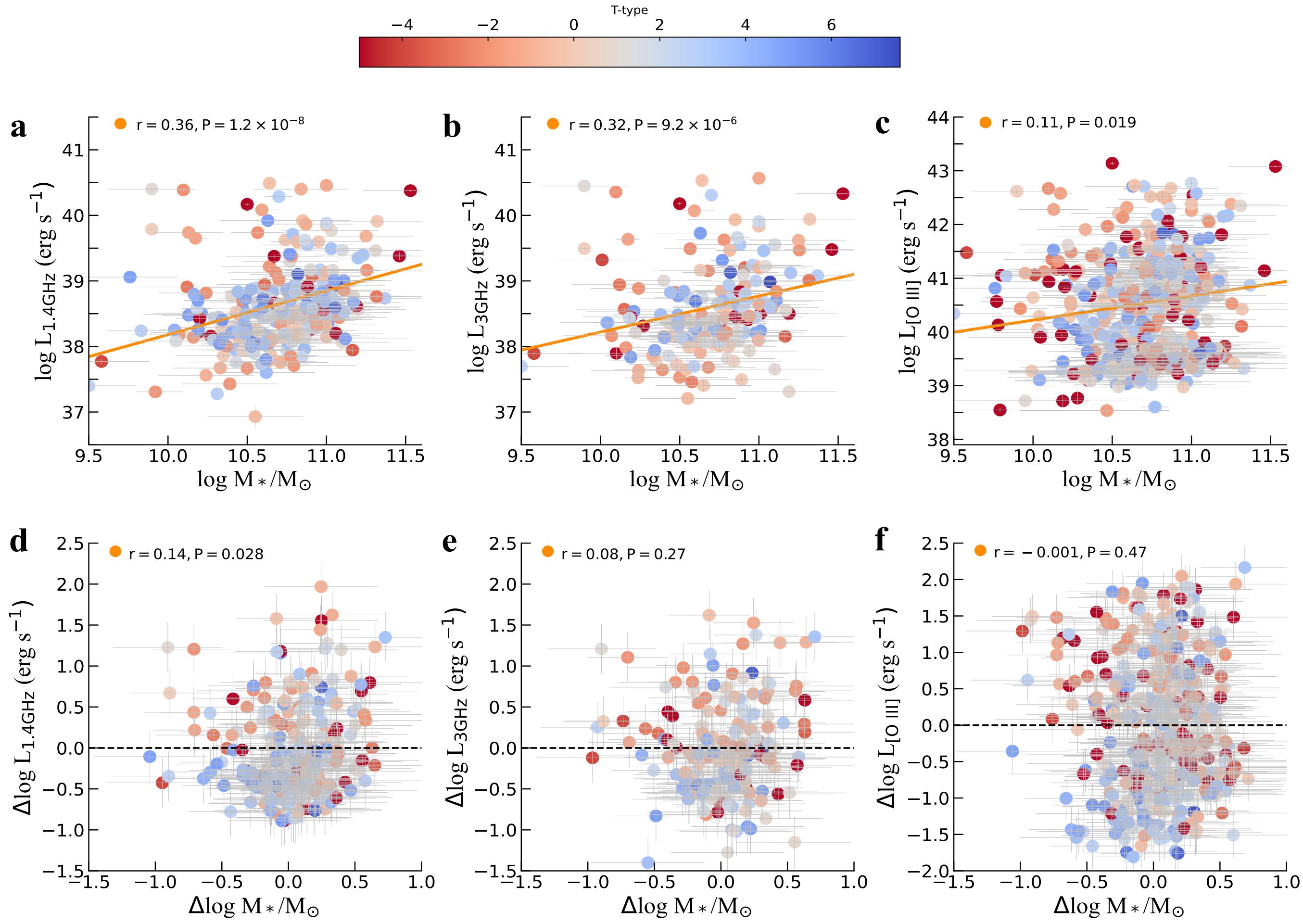

**Figure 4.** Relation between AGN luminosity and stellar mass for AGNs. (a), (b), (c): $L_{1.4\ \mathrm{GHz}}$–$M_*$ (a), $L_{3\ \mathrm{GHz}}$–$M_*$ (b), and $L_{[\mathrm{O\ III}]}$–$M_*$ (c) correlations. The color bar indicates morphological T types of host galaxies of AGNs (smaller values being more early-type and larger values more late-type galaxies). The orange lines are the best-fitted linear relation, taking into account the uncertainties of both variables. (d), (e), (f): comparison of the partial correlation of $L_{1.4\ \mathrm{GHz}}$–$M_*$ (while controlling for $M_{\mathrm{H_2}}$) (d), $L_{3\ \mathrm{GHz}}$–$M_*$ (while controlling for $M_{\mathrm{H_2}}$) (e), and $L_{[\mathrm{O\ III}]}$–$M_*$ (while controlling for $M_{\mathrm{H_2}}$) (f). The $x$- and $y$-axes are the residual in $M_*$, $L_{1.4\ \mathrm{GHz}}$, $L_{3\ \mathrm{GHz}}$, and $L_{[\mathrm{O\ III}]}$ after removing their dependence on $M_{\mathrm{H_2}}$ in the bottom panel: $\Delta \log Y = \log Y - \log Y(M_{\mathrm{H_2}})$ and $\Delta \log M_* = \log M_* - \log M_*(M_{\mathrm{H_2}})$.

L. C. Ho 2023). Therefore, we explore further the correlations between radio luminosity and [O III] luminosity and other main galactic parameters (A. Saintonge & B. Catinella 2022), including black hole mass ($\mathrm{M_{BH}}$), stellar mass ($M_*$), and bulge mass ($\mathrm{M_{bulge}}$), to ascertain whether molecular gas mass is a pivotal parameter determining AGN activity. The relationship between radio luminosity and [O III] luminosity and black hole mass is shown in Figure 3. We find that radio luminosity and [O III] luminosity have a weaker correlation with black hole mass than with molecular gas mass, with $r_{1.4\ \mathrm{GHz}} = 0.28$, $P = 5.1 \times 10^{-6}$; $r_{3\ \mathrm{GHz}} = 0.17$, $P = 0.02$, and $r_{[\mathrm{O\ III}]} = 0.18$, $P = 3.2 \times 10^{-4}$, respectively. The partial correlation analysis reveals that those correlations diminish when controlling for molecular gas mass ($M_{\mathrm{H_2}}$), resulting in reduced correlation coefficients: $r_{1.4\ \mathrm{GHz}} = 0.24$, $P = 1.1 \times 10^{-4}$; $r_{3\ \mathrm{GHz}} = 0.14$, $P = 0.06$, and $r_{[\mathrm{O\ III}]} = 0.13$, $P = 0.008$. In contrast, the correlations between radio luminosity, [O III] luminosity, and molecular gas mass remain robust even when controlling for black hole mass and stellar mass.

The correlations between radio luminosity and [O III] luminosity and stellar mass are shown in Figure 4. Radio luminosities and [O III] luminosity also show weaker correlations with stellar mass than with molecular gas mass, with respective correlation coefficients of $r_{1.4\ \mathrm{GHz}} = 0.36$, $P = 1.2 \times 10^{-8}$; $r_{3\ \mathrm{GHz}} = 0.32$, $P = 9.2 \times 10^{-6}$, and $r_{[\mathrm{O\ III}]} = 0.11$, $P = 0.019$. A partial correlation analysis shows that the correlations between 3 GHz radio luminosity, [O III] luminosity, and stellar mass nearly vanish when controlling for molecular gas mass, yielding correlation coefficients of $r_{1.4\ \mathrm{GHz}} = 0.14$, $P = 0.028$; $r_{3\ \mathrm{GHz}} = 0.08$, $P = 0.27$, and $r_{[\mathrm{O\ III}]} = -0.001$, $P = 0.47$.

Furthermore, there is a significant correlation between radio luminosity and bulge mass for AGNs (see Figure 5), with $r_{1.4\ \mathrm{GHz}} = 0.35$, $P = 0.098$ and $r_{3\ \mathrm{GHz}} = 0.50$, $P = 3.8 \times 10^{-4}$, although this correlation is less pronounced for [O III] luminosity ($r_{[\mathrm{O\ III}]} = 0.21$, $P = 0.011$). Notably, the correlations between radio luminosity and [O III] luminosity and bulge mass become weaker when molecular gas mass is excluded from the analysis, with correlation coefficients of $r_{1.4\ \mathrm{GHz}} = 0.19$, $P = 0.15$; $r_{3\ \mathrm{GHz}} = 0.33$, $P = 0.02$, and $r_{[\mathrm{O\ III}]} = 0.13$, $P = 0.127$.

Considering the high correlation among $M_{\mathrm{H_2}}$, $M_{\mathrm{BH}}$, $M_*$, and $\mathrm{M_{bulge}}$, we use a partial least-squares regression analysis on the AGN sample to further investigate the fundamental role of $\mathrm{M_{H_2}}$ in driving the correlations with radio luminosity and [O III] luminosity. The analysis included radio luminosity, [O III] luminosity, and the parameter sets $M_{\mathrm{H_2}}$, $M_{\mathrm{BH}}$, $M_*$, and $M_{\mathrm{bulge}}$. Our results indicate that $M_{\mathrm{H_2}}$ is the most significant predictor for both radio luminosity and [O III] luminosity (see Table 3). This suggests that, of all the properties we have considered, it is that the molecular gas mass may thus be the most important driver of nuclear activity.

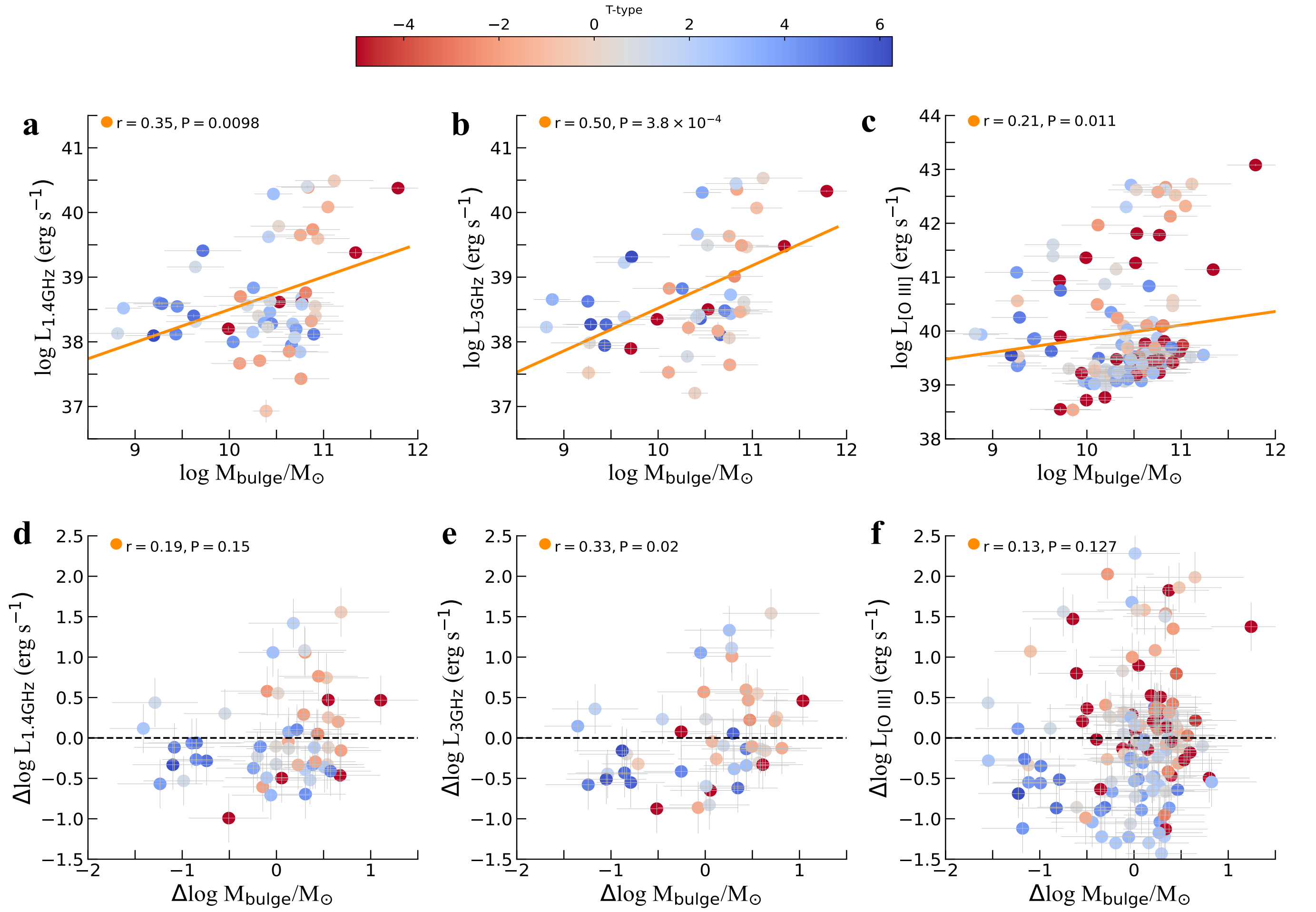

**Figure 5.** Relation between AGN luminosity and bulge mass for AGNs. (a), (b), (c): $L_{1.4\ \mathrm{GHz}}$–$M_{\mathrm{bulge}}$ (a), $L_{3\ \mathrm{GHz}}$–$M_{\mathrm{bulge}}$ (b), and $L_{[\mathrm{O\ III}]}$–$M_{\mathrm{bulge}}$ (c) correlations. The color bar indicates morphological T types of host galaxies of AGNs (smaller values being more early-type and larger values more late-type galaxies). The orange lines are the best-fitted linear relation, taking into account the uncertainties of both variables. (d), (e), (f): comparison of the partial correlation of $L_{1.4\ \mathrm{GHz}}$–$M_{\mathrm{bulge}}$ (while controlling for $M_{\mathrm{H_2}}$) (d), $L_{3\ \mathrm{GHz}}$–$M_{\mathrm{bulge}}$ (while controlling for $M_{\mathrm{H_2}}$) (e), and $L_{[\mathrm{O\ III}]}$–$M_{\mathrm{bulge}}$ (while controlling for $M_{\mathrm{H_2}}$) (f). The $x$- and $y$-axes are the residual in $M_{\mathrm{bulge}}$, $L_{1.4\ \mathrm{GHz}}$, $L_{3\ \mathrm{GHz}}$, and $L_{[\mathrm{O\ III}]}$ after removing their dependence on $M_{\mathrm{H_2}}$ in the bottom panel: $\Delta \log Y = \log Y - \log Y(M_{\mathrm{H_2}})$ and $\Delta \log M_* = \log M_* - \log M_*(M_{\mathrm{H_2}})$.

In the study of galaxy formation, various models of gas accretion onto black holes have been utilized. For instance, the Bondi accretion model has been applied in Illustris simulations (D. Sijacki et al. 2015). Additionally, some semianalytical models of galaxy formation incorporate an empirical model of cold gas accretion (M. Enoki et al. 2014). Our observation results are consistent with the numerical simulations.

### *3.3. Relation between Star Formation Rates and AGN Power*

In the Universe, AGN feedback regulates star formation in massive galaxies. It is anticipated that there is a robust correlation between star formation and AGN luminosity. If the gas driving star formation also fuels the central black hole, a positive correlation would be expected (T. A. Gutcke et al. 2015). Therefore, we study the relationship between SFR and the luminosity of AGNs. The relationships between radio luminosity and [O III] luminosity and SFRs are shown in Figure 6. A significant correlation persists between radio luminosity, [O III] luminosity, and SFR, even after accounting for the influence of black hole mass ($r_{1.4\ \mathrm{GHz}} = 0.64$, $P = 1.07 \times 10^{-28}$;$r_{3\ \mathrm{GHz}} = 0.65$, $P = 8.32 \times 10^{-23}$ and $r_{[\mathrm{O\ III}]} = 0.50$, $P = 3.17 \times 10^{-27}$; in middle panels of in Figure 6) and stellar mass ($r_{1.4\ \mathrm{GHz}} = 0.56$, $P = 1.68 \times 10^{-21}$;$r_{3\ \mathrm{GHz}} = 0.57$, $P = 8.32 \times 10^{-17}$ and $r_{[\mathrm{O\ III}]} = 0.45$, $P = 5.17 \times 10^{-22}$; in bottom panels of in Figure 6). Meanwhile, we also find a significant correlation between 5100 Å luminosity and SFRs ($r = 0.56$, $P = 5.62 \times 10^{-32}$), molecular gas mass ($r = 0.45$, $P = 7.4 \times 10^{-20}$), and molecular gas fraction ($r = 0.44$, $P = 1.11 \times 10^{-19}$) for AGNs (see Figure 7). These correlations between AGN luminosity and SFR can be interpreted as the evidence for the synchronized growth of black holes and their host galaxies. This suggests that the existing cold gas supply on the galactic scale can simultaneously provide fuel for the central black hole and facilitate significant star formation in massive galaxies.

Previous studies using aromatic features to evaluate the level of star formation in AGN host galaxies have found that it is related to AGN luminosity (M. Imanishi & K. Wada 2004; M. Schweitzer et al. 2006; H. Netzer et al. 2007; D. Lutz et al. 2008). These findings are broadly consistent with models where both star formation and AGN activity are triggered by an external supply of cold gas, or where nuclear star formation fuels subsequent AGN activity. Some authors used the

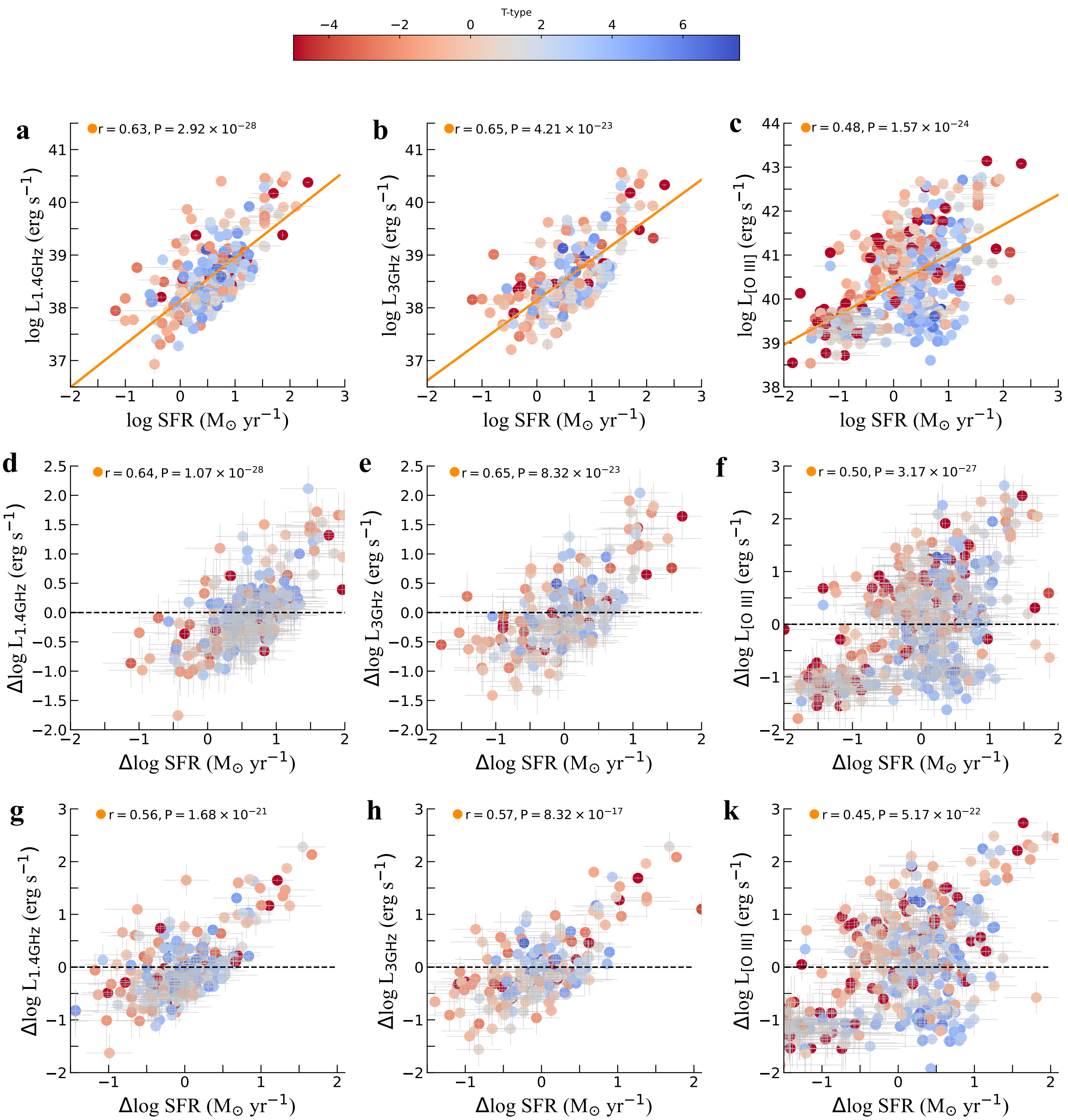

**Figure 6.** Relation between AGN luminosity and star formation rates for AGNs. (a), (b), (c): $L_{1.4\ \mathrm{GHz}}$–SFR (a), $L_{3\ \mathrm{GHz}}$–SFR (b), and $L_{[\mathrm{O\ III}]}$–SFR (c) correlations. The color bar indicates morphological T types of host galaxies of AGNs (smaller values being more early-type and larger values more late-type galaxies). The orange lines are the best-fitted linear relation, taking into account the uncertainties of both variables. (d), (e), (f), (g), (h), (k): comparison of the partial correlation of $L_{1.4\ \mathrm{GHz}}$–SFR (while controlling for $M_{\mathrm{BH}}$) (d), $L_{3\ \mathrm{GHz}}$–SFR (while controlling for $M_{\mathrm{BH}}$) (e), and $L_{[\mathrm{O\ III}]}$–SFR (while controlling for $M_{\mathrm{BH}}$) (f), $L_{1.4\ \mathrm{GHz}}$–SFR (while controlling for $M_*$) (g), $L_{3\ \mathrm{GHz}}$–SFR (while controlling for $M_*$) (h), and $L_{[\mathrm{O\ III}]}$–SFR (while controlling for $M_*$) (k). The $x$- and $y$-axes are the residual in SFR, $L_{1.4\ \mathrm{GHz}}$, $L_{3\ \mathrm{GHz}}$, and $L_{[\mathrm{O\ III}]}$ after removing their dependence on $M_{\mathrm{BH}}$ in the middle panel, and after removing their dependence on $M_*$ in the bottom panel: $\Delta \log Y = \log Y - \log Y(M_{\mathrm{BH}})$ and $\Delta\mathrm{logSFR} = \mathrm{logSFR} - \mathrm{logSFR}(M_{\mathrm{BH}})$ in the middle panel, and $\Delta \log Y = \log Y - \log Y(M_*)$ and $\Delta\mathrm{logSFR} = \mathrm{logSFR} - \mathrm{logSFR}(M_*)$ in the bottom panel.

emission lines (e.g., [O II], [O III], [Ne II], [Ne III]) to estimate the SFR and found a significant correlation between black hole accretion rate and SFRs (M.-Y. Zhuang & L. C. Ho 2020; M.-Y. Zhuang et al. 2021; M.-Y. Zhuang & L. C. Ho 2022). These correlations can be explained as evidence that star formation is influenced by positive feedback from AGNs. M.-Y. Zhuang et al. (2021) suggested that the correlation between SFRs and black hole accretion rate depends on the molecular gas mass; the correlation between SFRs and black hole accretion rate is still weak after removing the molecular gas mass. M.-Y. Zhuang et al. (2021) further showed that the star formation efficiency is correlated with black hole accretion

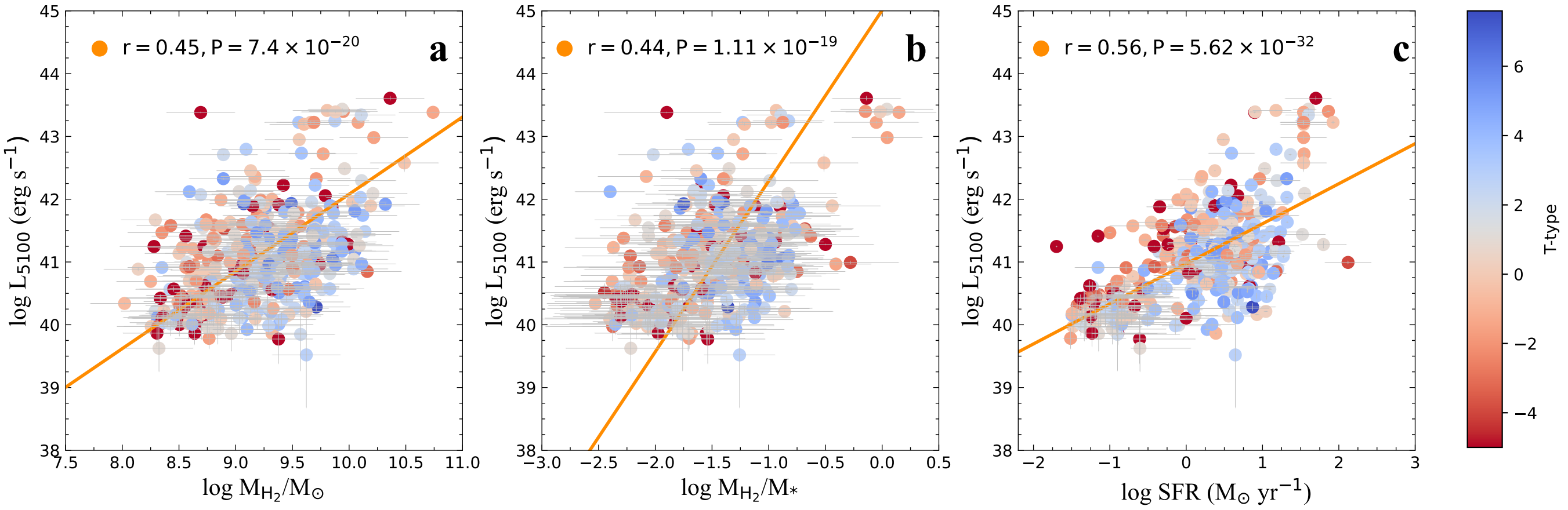

**Figure 7.** Relation between 5100 Å luminosity and star formation rates and molecular gas for AGNs. (a), (b), (c): $L_{5100}$–$M_{\rm H_2}$(a), $L_{5100}$–$M_{\rm H_2}/M_*$ (b), and $L_{5100}$–SFR (c) correlations. The color bar indicates morphological T types of host galaxies of AGNs (smaller values being more early-type and larger values more late-type galaxies). The orange lines are the best-fitted linear relation, taking into account the uncertainties of both variables.

**Table 3**
Results of Partial Correlation Analysis for AGNs (Both Early-type and Late-type Galaxies)

| Variables | | | | Correlation | |
|---|---|---|---|---|---|
| $X_1$ | $X_2$ | $X_3$ | $X_4$ | $r$ | $P$ |
| $\log L_{1.4\,\rm GHz}$ | $\log M_{\rm H_2}$ | $\log M_{\rm BH}$ | $\log M_*$ | 0.49 | $7.31 \times 10^{-16}$ |
| $\log L_{3\,\rm GHz}$ | $\log M_{\rm H_2}$ | $\log M_{\rm BH}$ | $\log M_*$ | 0.54 | $1.18 \times 10^{-14}$ |
| $\log L_{\rm [O\,III]}$ | $\log M_{\rm H_2}$ | $\log M_{\rm BH}$ | $\log M_*$ | 0.35 | $6.44 \times 10^{-13}$ |
| $\log L_{5100\,\rm Å}$ | $\log M_{\rm H_2}$ | $\log M_{\rm BH}$ | $\log M_*$ | 0.46 | $1.34 \times 10^{-20}$ |
| $\log L_{1.4\,\rm GHz}$ | $\log M_{\rm H_2}/M_*$ | $\log M_{\rm BH}$ | $\log M_*$ | 0.44 | $1.26 \times 10^{-12}$ |
| $\log L_{3\,\rm GHz}$ | $\log M_{\rm H_2}/M_*$ | $\log M_{\rm BH}$ | $\log M_*$ | 0.51 | $5.36 \times 10^{-13}$ |
| $\log L_{\rm [O\,III]}$ | $\log M_{\rm H_2}/M_*$ | $\log M_{\rm BH}$ | $\log M_*$ | 0.35 | $3.32 \times 10^{-13}$ |
| $\log L_{5100\,\rm Å}$ | $\log M_{\rm H_2}/M_*$ | $\log M_{\rm BH}$ | $\log M_*$ | 0.46 | $1.66 \times 10^{-20}$ |
| $\log L_{1.4\,\rm GHz}$ | $\log M_{\rm BH}$ | $\log M_{\rm H_2}$ | $\log M_*$ | 0.19 | 0.002 |
| $\log L_{3\,\rm GHz}$ | $\log M_{\rm BH}$ | $\log M_{\rm H_2}$ | $\log M_*$ | 0.06 | 0.38 |
| $\log L_{\rm [O\,III]}$ | $\log M_{\rm BH}$ | $\log M_{\rm H_2}$ | $\log M_*$ | 0.15 | 0.001 |
| $\log L_{5100\,\rm Å}$ | $\log M_{\rm BH}$ | $\log M_{\rm H_2}$ | $\log M_*$ | −0.06 | 0.19 |
| $\log L_{1.4\,\rm GHz}$ | $\log M_*$ | $\log M_{\rm H_2}$ | $\log M_{\rm BH}$ | 0.11 | 0.09 |
| $\log L_{3\,\rm GHz}$ | $\log M_*$ | $\log M_{\rm H_2}$ | $\log M_{\rm BH}$ | 0.12 | 0.10 |
| $\log L_{\rm [O\,III]}$ | $\log M_*$ | $\log M_{\rm H_2}$ | $\log M_{\rm BH}$ | −0.07 | 0.14 |
| $\log L_{5100\,\rm Å}$ | $\log M_*$ | $\log M_{\rm H_2}$ | $\log M_{\rm BH}$ | −0.06 | 0.18 |
| $\log L_{1.4\,\rm GHz}$ | logSFR | $\log M_*$ | $\log M_{\rm BH}$ | 0.61 | $5.26 \times 10^{-26}$ |
| $\log L_{3\,\rm GHz}$ | logSFR | $\log M_*$ | $\log M_{\rm BH}$ | 0.61 | $6.03 \times 10^{-20}$ |
| $\log L_{\rm [O\,III]}$ | logSFR | $\log M_*$ | $\log M_{\rm BH}$ | 0.49 | $2.58 \times 10^{-26}$ |
| $\log L_{5100\,\rm Å}$ | logSFR | $\log M_*$ | $\log M_{\rm BH}$ | 0.55 | $3.08 \times 10^{-31}$ |

**Note.** $X_1$ and $X_2$ are variables that have a possible mutual dependency on variable $X_3$ and $X_4$. $r$ and $P$ are the Spearman partial correlation coefficient and probability for the null hypothesis of no correlation between $X_1$ and $X_2$, respectively.

rate, potentially indicative of positive AGN feedback. Y. Chen et al. (2025a) found a significant correlation between black hole spin and star formation in massive star-forming galaxies. Y. Chen et al. (2025b) found a significant correlation between the magnetic field of jets and SFR for a large sample of 96 galaxies hosting supermassive black holes. According to the theory of jet formation, the activity of jets is related to the spin and magnetic field of black holes (R. D. Blandford & R. L. Znajek 1977; R. D. Blandford & D. G. Payne 1982). Meanwhile, we also find a significant correlation between the luminosity of AGNs and SFRs. Therefore, the conclusions of Y. Chen et al. (2025a) and Y. Chen et al. (2025b) and our results imply that black hole activity might have a positive influence on star formation, namely, the positive feedback of AGNs. We also use partial correlation to analyze the relationship between radio luminosity, [O III] luminosity, 5100 Å luminosity, and SFR. There has always been strong correlation between radio luminosity, [O III] luminosity, 5100 Å luminosity, and SFR after removing the molecular gas mass ($r_{1.4\ \rm GHz} = 0.43$, $P = 2.12 \times 10^{-12}$; $r_{3\ \rm GHz} = 0.40$, $P = 2.45 \times 10^{-8}$; $r_{\rm [O\ III]} = 0.34$, $P = 2.11 \times 10^{-12}$; $r_{5100\ \rm Å} = 0.38$, $P = 6.69 \times 10^{-14}$).

The slopes of the relationships between radio luminosity and [O III] luminosity and molecular gas mass are 0.94 ± 0.10, 1.18 ± 0.14, and 0.96 ± 0.12 for our sample. The slopes of the relation between radio luminosity and [O III] luminosity and SFR are 0.82 ± 0.06, 0.76 ± 0.06, and 0.68 ± 0.05 for our sample (see Table 2). According to the predictions of the theoretical model, R. C. Hickox et al. (2014) found that the AGN luminosity is proportional to total far-IR luminosities ($L_{\rm IR}$, integrated from 8 to 1000 $\mu$m), $L_{\rm AGNs} \propto L_{\rm IR}^{1.0}$. The $\rm L_{IR}$ is also proportional to SFRs (R. C. Kennicutt 1998a), $L_{\rm IR} \propto {\rm SFR}^{1.0}$. Additionally, star formation in galaxies follows the Kennicutt–Schmidt law (R. C. Kennicutt 1998b), where the surface density of star formation ($\Sigma_{\rm SFR}$) is proportional to the surface density of molecular gas ($\Sigma_{\rm H_2}$), $\Sigma_{\rm SFR} \propto \Sigma_{\rm H_2}^{1.0}$ for AGN host galaxies (J. Shangguan et al. 2020a), namely, $SFR \propto M_{\rm H_2}^{1.0}$. Therefore, we have $L_{\rm AGNs} \propto SFR^{1.0} \propto M_{H_2}^{1.0}$. The slope of the relation between 1.4 GHz radio luminosity and SFRs for our sample is 0.82 ± 0.06, which is close to 1.0. Our results are consistent with the theoretical model proposed by R. C. Hickox et al. (2014).

At the same time, we also study the relationship between AGN activity and molecular gases in early- and late-type galaxies. We find that the slope of the relationship between radio luminosity, [O III] luminosity, 5100 Å luminosity, and molecular gas mass (molecular gas content) is slightly different for early-type and late-type galaxies. The correlation between the activity of AGNs and molecular gas is stronger in early-type galaxies than in late-type galaxies (see Table 2). Those results may suggest that the feedback from AGN activity in early-type galaxies is stronger than that in late-type galaxies.

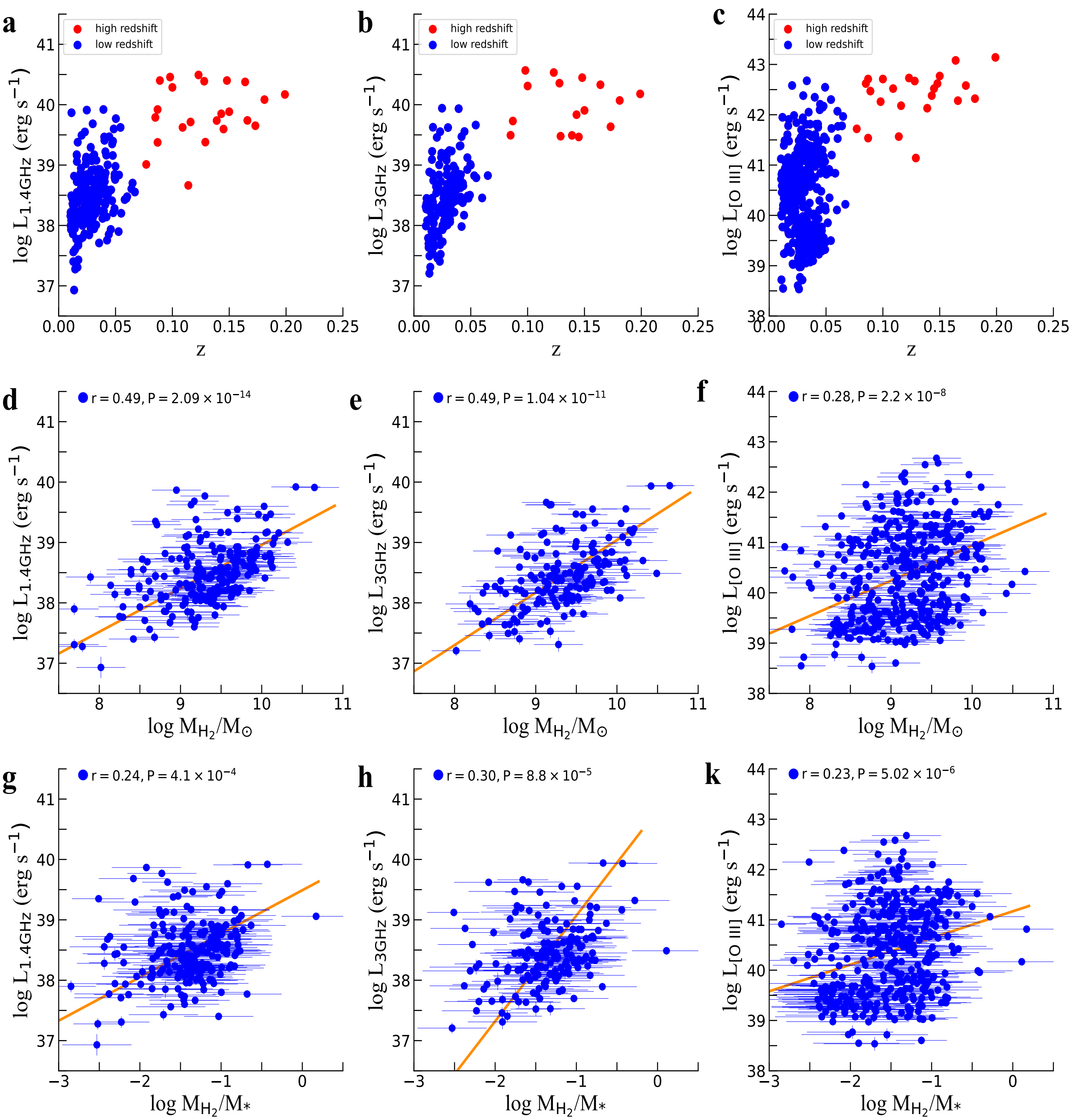

**Figure 8.** Relation between AGN luminosity and redshift (top panel), molecular gases (middle panel), and molecular gases content (bottom panel) for AGNs. (a), (b), (c): $L_{1.4\ \mathrm{GHz}}$–$z$ (a), $L_{3\ \mathrm{GHz}}$–$z$ (b), and $L_{[\mathrm{O\ III}]}$–$z$ (c) correlations. The orange lines are the best-fitted linear relation, taking into account the uncertainties of both variables. (d), (e), (f), (g), (h), (k): $L_{1.4\ \mathrm{GHz}}$–$M_{\mathrm{H_2}}$ (d), $L_{3\ \mathrm{GHz}}$–$M_{\mathrm{H_2}}$ (e), and $L_{[\mathrm{O\ III}]}$–$M_{\mathrm{H_2}}$ (f), $L_{1.4\ \mathrm{GHz}}$–$M_{\mathrm{H_2}}/M_*$ (g), $L_{3\ \mathrm{GHz}}$–$M_{\mathrm{H_2}}/M_*$ (h), and $L_{[\mathrm{O\ III}]}$–$M_{\mathrm{H_2}}/M_*$ (k). The red dot is high-redshift sources, and blue dot is low-redshift sources.

The relationship between luminosity and redshift for our sample is presented in Figure 8. A distinct population of high-redshift, high-luminosity AGNs forms a long tail that is separated from the majority of the sample—these sources are marked as red dots in the top panels of Figure 8. We recognize that the Malmquist bias may influence observed correlations between variables; therefore, we reassessed the correlations among radio luminosity, [O III] luminosity, 5100 Å luminosity, and molecular gas content, and molecular gas mass while accounting for this bias. To minimize its impact, we focus our analysis on low-redshift sources (represented by blue dots). Our results reveal significant correlations between radio luminosity (at both 1.4 and 3 GHz), [O III] luminosity, 5100 Å luminosity, and molecular gas mass for low redshift ($r_{1.4\ \mathrm{GHz}} = 0.49$, $P = 2.09 \times 10^{-14}$; $r_{3\ \mathrm{GHz}} = 0.49$, $P = 1.04 \times 10^{-11}$; $r_{[\mathrm{O\ III}]} = 0.28$, $P = 2.2 \times 10^{-8}$; $r_{5100\ \mathrm{Å}} = 0.38$, $P = 5.5 \times 10^{-14}$). Furthermore, strong correlations persist between radio luminosity, [O III] luminosity, and molecular gas content for low redshift ($r_{1.4\ \mathrm{GHz}} = 0.24$, $P = 4.1 \times 10^{-4}$; $r_{3\ \mathrm{GHz}} = 0.30$,

$P = 8.8 \times 10^{-3}$; $r_{[\text{O III}]} = 0.23$, $P = 5.02 \times 10^{-6}$; $r_{5100\,\text{Å}} = 0.39$, $P = 3.93 \times 10^{-14}$). These findings confirm that the main conclusions remain robust even after correcting for Malmquist bias, supporting the interpretation that molecular gas plays a key role in regulating AGN activity.

## 4. Conclusion

Cold molecular gases have been observed in massive galaxies, which likely provide fuel for their AGNs. To confirm this speculation, we collect large samples of AGNs with molecular gas mass and discussed their relationship to AGN activity. Our main results are as follows:

1. We find that radio luminosity and [O III] luminosity exhibit significant correlations with both the molecular gas mass and molecular gas content, even when the effects of black hole mass and stellar mass are excluded. There is also a significant correlation between 5100 Å luminosity and molecular gas mass and molecular gas content for our sample. These results suggest that AGNs are powered by molecular gas.
2. The correlations between radio luminosity and [O III] luminosity with black hole mass, stellar mass, and bulge mass become weaker when molecular gas mass is excluded. This suggests that, of all the properties we have considered, it is the molecular gas mass that is most tightly correlated with radio and [O III] luminosity, and may thus be the most important driver of nuclear activity.
3. The radio luminosity and [O III] luminosity show significant correlations with the SFR, even after accounting for the effects of black hole mass and stellar mass. There is also a significant correlation between 5100 Å luminosity and SFRs for our sample. These results imply that the activity of AGNs might have a positive influence on star formation.

## Acknowledgments

Y.C. is grateful for financial support from the National Natural Science Foundation of China (No. 12203028). Y.C. is grateful for funding for the training program for talent in Xingdian, Yunnan Province (2081450001). Q.S.G.U. is supported by the National Natural Science Foundation of China (12121003, 12192220, and 12192222). This work is supported by the National Natural Science Foundation of China (11733001, U2031201, and 12433004). X.G. acknowledges the support of National Nature Science Foundation of China (No. 12303017). This work is also supported by Anhui Provincial Natural Science Foundation project No. 2308085QA33. D.R.X. is supported by the NSFC 12473020, Yunnan Province Youth Top Talent Project (YNWR-QNBJ-2020-116), and the CAS Light of West China Program.

## ORCID iDs

Yongyun Chen (陈永云) https://orcid.org/0000-0001-5895-0189
Qiusheng Gu (顾秋生) https://orcid.org/0000-0002-3890-3729
Luis.C Ho (何子山) https://orcid.org/0000-0001-6947-5846
Junhui Fan (樊军辉) https://orcid.org/0000-0002-5929-0968
Feng Yuan (袁峰) https://orcid.org/0000-0003-3564-6437
Tao Wang (王涛) https://orcid.org/0000-0002-2504-2421
Zhifu Chen (陈志福) https://orcid.org/0000-0003-0639-1148
Dingrong Xiong (熊定荣) https://orcid.org/0000-0002-6809-9575
Xiaotong Guo (郭晓通) https://orcid.org/0000-0002-2338-7709
Nan Ding (丁楠) https://orcid.org/0000-0003-1028-8733